\documentclass[twocolumn]{aastex6}
\usepackage{amsmath,amstext, bm}
\usepackage{amssymb}
\usepackage{epsfig}
\usepackage{color}
\usepackage{hyperref}
\usepackage{float}

\begin{document}

\title{A two-zone model for blazar emission: implications for TXS~0506+056 and the neutrino event IceCube-170922A}

\author{Rui Xue\altaffilmark{1, 7}, Ruo-Yu Liu\altaffilmark{2}, Maria Petropoulou\altaffilmark{3}, Foteini Oikonomou\altaffilmark{4}, Ze-Rui Wang\altaffilmark{1, 7},  Kai Wang\altaffilmark{5, 6} and Xiang-Yu Wang\altaffilmark{1, 7}}
\altaffiltext{1}{School of Astronomy and Space Science, Nanjing University, Nanjing 0110093, China}
\altaffiltext{2}{Deutsches Elektronen Synchrotron (DESY), Platanenallee 6, D-15738 Zeuthen, Germany}
\altaffiltext{3}{Department of Astrophysical Sciences, Princeton University, 4 Ivy Lane, Princeton, NJ 08544, USA}
\altaffiltext{4}{European Southern Observatory, Karl-Schwarzschild-Str 2, D-85748 Garching bei Munchen, Germany}
\altaffiltext{5}{Department of Astronomy, School of Physics, Peking University, Beijing 100871, China}
\altaffiltext{6}{Kavli Institute for Astronomy and Astrophysics, Peking University, Beijing 100871, China}
\altaffiltext{7}{Key laboratory of Modern Astronomy and Astrophysics (Nanjing University), Ministry of Education, Nanjing 210023, People's Republic of China }

\begin{abstract}
A high-energy muon neutrino event, IceCube-170922A, was recently discovered in both spatial and temporal coincidence with a gamma-ray flare of the blazar TXS~0506+056.
It has been shown, with standard one-zone models, that neutrinos can be produced in the blazar jet via hadronic interactions, but with a flux which is mostly limited by the X-ray data.
In this work, we explore the neutrino production from TXS~0506+056 by invoking two physically distinct emission zones in the jet, separated by the broad line region (BLR). Using the Doppler-boosted radiation of the BLR as the target photon field, the inner zone accounts for the neutrino and gamma-ray emission via $p\gamma$ interactions and inverse Compton scattering respectively, while the outer zone produces the optical and X-ray emission via synchrotron and  synchrotron self-Compton processes. The different conditions of the two zones allow us to suppress the X-ray emission from the electromagnetic cascade, and set a much higher upper limit on the muon neutrino flux (i.e., $\sim 10^{-11}\rm erg~cm^{-2}s^{-1}$) than in one-zone models. We compare, in detail, our scenario with one-zone models discussed in the literature, and argue that differentiating between such scenarios will become possible with next generation neutrino telescopes, such as IceCube-Gen2.
\end{abstract}



\section{Introduction}\label{intro}
Blazars are a special class of active galactic nuclei (AGN), with relativistic jets closely aligned to our line of sight \citep{1995PASP..107..803U}. Being some of the most powerful persistent sources of electromagnetic (EM) radiation in the Universe, with typical bolometric luminosities of $10^{43}$-$10^{48}$~erg s$^{-1}$, blazars have been widely considered as candidate sources for high-energy cosmic rays and neutrinos \citep[e.g.,][]{1992A&A...260L...1M, 2001PhRvL..87v1102A, 2014PhRvD..90b3007M, 2015MNRAS.448.2412P, 2016MNRAS.457.3582P, 2019ApJ...871...41P}.

On September 22, 2017, a 290 TeV muon neutrino, IceCube-170922A, triggered the IceCube extremely-high-energy (EHE) online event filter \citep{2018Sci...361.1378I}. For the first time, a high-energy neutrino was detected in temporal and spatial coincidence with a multi-wavelength flare from a known blazar (TXS~0506+056), with a significance of $\sim3\sigma$. This event has been extensively studied in the framework of  conventional one-zone leptonic models with photohadronic ($p\gamma$) interactions \citep{2018ApJ...863L..10A,2018ApJ...864...84K,2019NatAs...3...88G,2019MNRAS.483L..12C}. In these models, the multi-wavelength spectral energy distribution (SED) of the flare is explained by the synchrotron radiation and Compton processes
of relativistic electrons injected into a localized radiation zone  inside the jet (the so-called emitting blob), while neutrinos are produced via the $p\gamma$ interaction between relativistic protons and synchrotron photons. 
Electron-positrons pairs (henceforth, secondary pairs) are also produced by the $p\gamma$ interactions and Bethe-Heitler (BH) pair production processes of relativistic protons. Extremely high-energy photons, produced either by the decay of neutral pions or by the synchrotron emission of secondary pairs, are absorbed in the soft (synchrotron) radiation field via  $\gamma\gamma$ pair production \citep[e.g.,][]{1992A&A...253L..21M}, thus transferring their energy to relatively lower-energy photons (e.g., from X-rays to MeV and TeV $\gamma$-rays) through an EM cascade in the blob \citep[e.g.,][]{2015MNRAS.447...36P}. The (quasi-)simultaneous X-ray observations from {\it Swift}  and {\it NuSTAR}  provide a stringent constraint on the cascade emission \citep{2018ApJ...864...84K,2018ApJ...865..124M}. Since both the EM cascade emission and the neutrino emission originate from the $p\gamma$ interactions on the same target photon field, the X-ray data can also constrain the neutrino flux, yielding a detection rate $<0.03 \, \rm yr^{-1}$ \citep{2018ApJ...864...84K, 2019NatAs...3...88G} after convolving with the effective area of the IceCube EHE track-event alert system\footnote{In some literature, the neutrino detection rate is calculated with the normal IceCube effective area for muon neutrino which is about 10 times higher than that of the EHE through-going muon-track event filter, resulting in a 10 times higher detection rate than the one presented here.} \citep{2018Sci...361.1378I}. 


Such a low detection rate translates to a probability of $\lesssim 1\%$ for IC-170922A to have triggered IceCube's alert system, assuming that the blazar flare lasted only several months. While it may be possible to explain the detection of one neutrino from TXS\,0506+056 
as a consequence of the Eddington bias \citep{2019A&A...622L...9S}, it is worth investigating if 
a higher neutrino flux and hence a more comfortable detection probability of the event can be obtained in other theoretical scenarios. \cite{2019PhRvD..99f3008L} proposed that a high neutrino flux can be obtained, if a second emitting zone forms inside the broad-line region (BLR) of the blazar, in addition to the emitting blob considered in the conventional one-zone model which accounts for the EM emission of the blazar. In that scenario, accelerated protons interact with BLR clouds that enter the jet and produce neutrinos efficiently via proton--proton ($pp$) inelastic collisions, with the resulting X-ray flux still being consistent with the observations. The idea of two emitting zones and their potential to improve the predicted neutrino detection rate has also been discussed by \cite{2018ApJ...865..124M} in the context of the neutral beam model \citep{2003ApJ...586...79A, 2012ApJ...755..147D} and by \cite{2019ApJ...876..109Z}. 

Generally speaking, two-zone scenarios invoke a second emitting region with additional free parameters, thus introducing more uncertainties to the model. However, the presence of multiple emission sites in blazar jets is physically plausible. For example, 
substructures, such as multiple bright knots and hot spots, have been observed along the jets of radio galaxies \citep[see e.g.,][for a review]{2006ARA&A..44..463H}, which are thought to be the off-axis counterparts of blazars \citep{1995PASP..107..803U,2010ApJ...716...30A}, implying that particle acceleration and emission can take place at different parts of a jet at the same time. These features have been discovered in jets of different lengths, spanning from kiloparsec to Megaparsec scales. Here, we speculate that multiple radiation zones may also exist in the inner (sub-)parsec scale jets of blazars. The recent discovery of significant offsets between the optical core and the radio core pinpointed respectively with Gaia and VLBI in many blazars \citep{2019ApJ...871..143P} may lend support to our assumption. There are also indications of multiple emitting zones in individual blazars \citep[e.g.,][]{2012ApJ...760...69N, 2019arXiv190804803P}.
Even if the intrinsic properties of different emitting zones, such as their Doppler factor and size, are similar to each other, their external environment may be different. 
For example, if the emitting region is located within the BLR, the latter provides an extra target photon field for the inverse Compton scattering of electrons and for the photohadronic ($p\gamma$) interactions of protons. It may also provide the dense clouds required by the $pp$ collision mechanism \citep{2018ApJ...866..109S, 2019PhRvD..99f3008L}. Therefore, the resulting radiation from an emitting zone inside (close to) the BLR can be very different from that outside (far away from) the BLR, when considering the properties of the external environment. 

TXS~0506+056 is classified as a 
BL Lac object based on the low equivalent width (EW) of three very weak emission lines \citep{stickel91, stocke91} identified in its optical spectrum \citep{paiano18}. However, this should not be interpreted as evidence for the lack of a BLR,  
since the radiation of the latter could be overwhelmed by the intense non-thermal emission from the jet \citep{1998ApJ...506..621G, 2012MNRAS.420.2899G, 2013MNRAS.431.1914G}. Interestingly, a recent paper by \citet{2019MNRAS.484L.104P} pointed out that TXS~0506+056 is in fact a masquerading BL Lac object, with a hidden BLR with luminosity  $\approx 5\times 10^{43}$~erg s$^{-1}$ and a standard accretion disk. 
\begin{figure*}[htbp]
\includegraphics[width=1\textwidth]{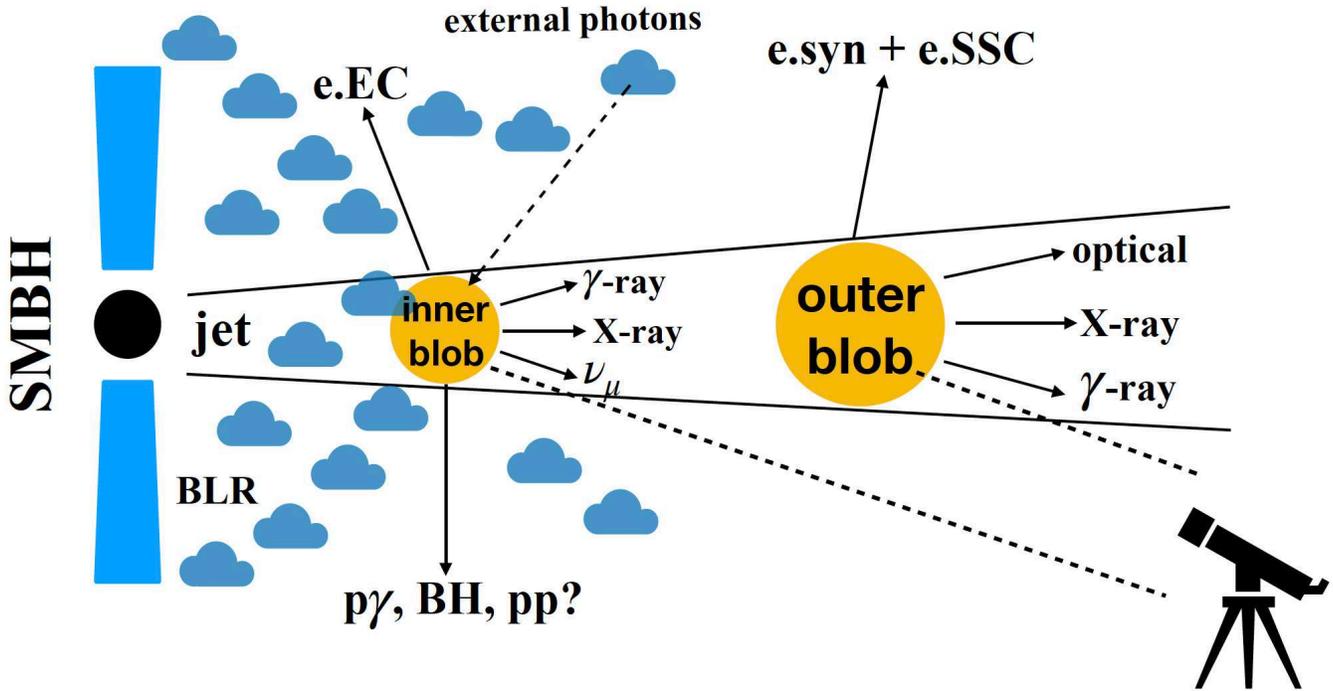}
\caption{A schematic illustration (not to scale) of the two-zone photohadronic ($p\gamma$) model. The outer blob is far away from the BLR and the inner blob is close or inside the BLR. In the presence of the external radiation field provided by the BLR, the inner blob is assumed to be an efficient $\gamma$-ray (via external Compton emission of electrons, denoted here as e.EC) and neutrino emitter due to the $p\gamma$ process or possibly the $pp$ process. The emission of the cascade initiated by electromagnetic secondaries produced in $p\gamma$, Bethe-Heitler or $pp$ processes can emerge in X-rays and gamma rays. The low-energy radiation is probably suppressed due to the dominant Compton cooling of electrons. The observed low-energy radiation arises from the synchrotron emission of electrons (e.syn) from the outer blob  where the BLR radiation is negligible. The outer blob may also produce hard X-ray and $\gamma$-ray photons via the SSC process (e.SSC) of accelerated electrons.
\label{fig:sketch}}
\end{figure*}

Motivated by both theoretical and observational perspectives, we set to explore the neutrino emission from  TXS~0506+056 in the framework of a two-zone $p\gamma$ model that takes into account the existence of the BLR. Our main goals are to compute the maximum neutrino flux predicted by the model, while fitting the SED with two radiation zones, and to understand the fundamental differences with the standard one-zone models.
The rest of this paper is structured as follows. In Section~\ref{model} we present the general description of the two-zone $p\gamma$ model and apply it to the 2017 observations of TXS~0506+056. In Section~\ref{compare} we show the differences between the two-zone $p\gamma$ model and the one-zone model, and explain the reasons why the former model can result in higher maximal neutrino fluxes. In Section~\ref{dis} we discuss our results, while paying particular attention to potential methods of differentiating among models in the future. Our conclusions are presented in Section~\ref{con}. Throughout the paper, we adopt the $\Lambda$CDM cosmology with $H_{\rm{0}}={70\rm{km~s^{-1} Mpc^{-1}}}$, $\Omega_{\rm{m}}=0.3$, $\Omega_{\rm{\Lambda}}=0.7$.

\section{Model description and application}\label{model}
Although there might be multiple emitting zones in the blazar jet,  here we consider the simplest multi-zone scenario which consists of two regions, one
far away from the BLR (the outer blob) and another one close to or inside the BLR (the inner blob). 
In the presence of the external radiation field provided by the BLR, the inner blob is assumed to be an efficient gamma-ray and neutrino emitter, while the low-energy (i.e., from infrared to soft X-ray) radiation of  synchrotron origin is likely suppressed. In our scenario, the observed low-energy radiation most likely arises from the outer blob where the BLR radiation is negligible. The outer blob may also produce hard X-ray and gamma-ray photons via the SSC process of electrons. A sketch of our model is shown in Fig.~\ref{fig:sketch}.

A detailed description of the model can be found in the following subsections.
{Henceforth, physical quantities with the superscript ``AGN'' are measured in the AGN frame, whereas quantities without the superscript are measured in the jet's comoving frame, unless otherwise specified.}



\subsection{The leptonic emission from the outer blob}\label{lep}
Relativistic electrons are assumed to be injected in the outer blob with a smooth broken power-law energy distribution \citep{2010MNRAS.402..497G} at a constant rate given by:
\begin{equation}\label{eq:einj}
\begin{split}
Q_{\rm e}(\gamma_{\rm e}) = &Q_{\rm e, 0}\gamma_{\rm e}^{-n_{\rm e, 1}}\left[1 + \left(\frac{\gamma_{\rm e}}{\gamma_{\rm e, b}}\right)^{(n_{\rm e, 2} - n_{\rm e, 1})}\right]^{-1}, \\
&{\rm for} ~~ \gamma_{\rm e, min} < \gamma_{\rm e} < \gamma_{\rm e, max},
\end{split}
\end{equation}
where $\gamma_{\rm e, min}$, $\gamma_{\rm e, b}$ and $\gamma_{\rm e, max}$ are the minimum, break and maximum electron Lorentz factors, $n_{\rm e, 1}$ and $n_{\rm e, 2}$ are the power-law indices below and above $\gamma_{\rm e, b}$, respectively, and $Q_{\rm e, 0}$ is a normalization constant. This is determined by
$\int Q_{\rm e}\gamma_{\rm e} m_{\rm e} c^2d\gamma_{\rm e}= 3\,L_{\rm e, inj}/(4 \pi R_{\rm out}^3)$, where $L_{\rm e, inj}$ is the electron injection luminosity and $R_{\rm out}$ is the radius of the outer blob, $m_{\rm e}$ is the electron rest mass, and $c$ is the speed of light. The steady-state electron density distribution can be then written as: 
\begin{equation}\label{eq:espec}
N_{\rm e}(\gamma_{\rm e}) \approx Q_{\rm e}(\gamma_{\rm e})t_{\rm e},
\end{equation}
where $t_{\rm e} = \min\{ t_{\rm cool}, t_{\rm esc, out}\}$ with  $t_{\rm esc, out} = R_{\rm out}/c$ being the particle escape timescale of the outer blob (in the ballistic propagation limit) and $t_{\rm cool} =  3m_{\rm e}c/(4(U_{\rm B}+\kappa_{\rm KN}U_{\rm ph})\sigma_{\rm T}\gamma_{\rm e})$ being the electron radiative cooling timescale. Here, $U_{\rm B} = B^2/8\pi$ is the energy density of the magnetic field, $U_{\rm ph}$ is the energy density of the soft photons, $\sigma_{\rm T}$ is the Thomson scattering cross section, and $\kappa_{\rm KN}$ is a numerical factor accounting for Klein-Nishina effects \citep{2005MNRAS.363..954M}. Using the steady-state electron distribution $N_{\rm e}(\gamma_{\rm e})$, we can then compute the synchrotron and synchrotron self-Compton (SSC) emissions \citep{2001A&A...367..809K}. Since we assume that the outer blob is far away from the BLR, the external Compton (EC) emission is not considered here. 

Relativistic protons are also assumed to be injected in the outer blob with a power-law energy distribution  and at a constant rate:
\begin{equation}\label{eq:pinj}
Q_{\rm p}(\gamma_{\rm p}) = Q_{\rm p, 0}\gamma_{\rm p}^{-n_{\rm p}}, \gamma_{\rm p, min} < \gamma_{\rm p} < \gamma_{\rm p, max},
\end{equation}
where $n_{\rm p}$ is the power-law index,  $\gamma_{\rm p, min}$ and $\gamma_{\rm p, max}$ are the minimum and maximum proton Lorentz factors, respectively, and $Q_{\rm p, 0}$ is a normalization constant.  This is determined by $\int Q_{\rm p}\gamma_{\rm p} m_{\rm p} c^2d\gamma_{\rm p}= 3\,L_{\rm p, inj}/(4 \pi R_{\rm out}^3)$, where $L_{\rm p, inj}$ is the proton injection luminosity and $m_{\rm p}$ is the proton rest mass. Similar to the electrons, the steady-state proton distribution can be written as: 
\begin{equation}\label{eq:pspec}
N_{\rm p}(\gamma_{\rm p}) \approx Q_{\rm p}(\gamma_{\rm p})t_{\rm p},
\end{equation}
where $t_{\rm p} = \min\{ t_{p\gamma}, t_{\rm BH}, t_{\rm esc,out}, t_{\rm p, syn}\}$.  More specifically, $t_{p\gamma} = [c n_{\rm soft}<\sigma_{p\gamma}\kappa_{p\gamma}>]^{-1}$ is the $p\gamma$ energy loss timescale, where $n_{\rm soft}$ is the number density of the soft photons 
and $<\sigma_{p\gamma}\kappa_{p\gamma}>\simeq10^{-28} \rm cm^2$ is the inelasticity-weighted $p\gamma$ interaction cross-section. $t_{\rm BH} = [c n_{\rm soft}<\sigma_{\rm BH}\kappa_{\rm BH}>]^{-1}$ is the BH pair-production cooling timescale, where $\sigma_{\rm BH}$ and $\kappa_{\rm BH}$  are, respectively, the cross section and inelasticity for the BH pair production process \citep{1992ApJ...400..181C}. Finally, $t_{\rm p, syn} = 6\pi m_{\rm e}c^2/(c\sigma_{\rm T}B^2\gamma_{\rm p})(m_{\rm p}/m_{\rm e})^3$ is the proton-synchrotron cooling timescale.

\begin{figure}[htbp]
\centering
\includegraphics[width=1\columnwidth]{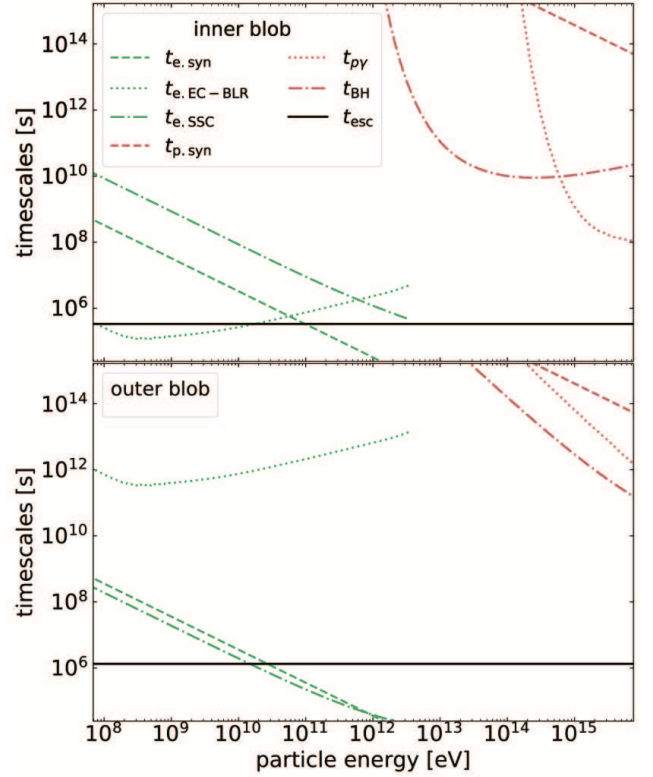}
\caption{Timescales of various cooling processes for electrons (green curves) and protons (red curves) in the inner blob (upper panel) and outer blob (lower panel) as a function of the particle energy. Both the particle energy and timescale are measured in the jet's comoving frame. Parameters are the same in those in Fig.~\ref{fig:sed}. The black horizontal lines denote the escape timescale of particles in the ballistic propagation limit.  The meaning of all curves is explained in the inset legend. \label{fig:timescale}}
\end{figure}

{In the outer blob, the only target radiation field for the $p\gamma$ process and the BH process is the synchrotron emission of primary electrons, which is inefficient for protons of energy $<10$~PeV (see Fig.~\ref{fig:timescale}). Thus, the hadronic emission of the outer blob can be neglected unless a highly super-Eddington proton kinetic luminosity is assumed.}

\subsection{The leptonic and hadronic emission from the inner blob}
We assume that electrons and protons are injected into the inner blob at the same rate as in the outer blob (see Eqs.~(\ref{eq:einj}) and (\ref{eq:pinj})). The steady-state energy spectra of primary electrons and protons are given by Eqs.~(\ref{eq:espec}) and (\ref{eq:pspec}), respectively, with the only difference that $U_{\rm ph}$ and $n_{\rm soft}$ are now dominated by the BLR photons due to the blob location in the jet (see also Fig.~\ref{fig:sketch}).

The energy density of BLR radiation in the jet comoving frame ($u_{\rm BLR}$) as a function of the {distance  along the jet, $r^{\rm AGN}_{\rm in}$, } can be approximately written as \citep{2012ApJ...754..114H}:
\begin{equation}\label{eq:ublr}
u_{\rm BLR} \approx \frac{\Gamma^2L_{\rm BLR}^{\rm AGN}}{4\pi (r_{\rm BLR}^{\rm AGN})^2c[1+(r^{\rm AGN}_{\rm in}/r_{\rm BLR}^{\rm AGN})^3]},
\end{equation}
where $\Gamma$ is the bulk Lorentz factor of the jet\footnote{For simplicity, we assume the Doppler factor $\delta_{\rm D} \approx \Gamma$ for a relativistic jet close to the line of sight in blazars with a viewing angle of $\theta \lesssim 1/\Gamma$ hereafter.}. {$L_{\rm BLR}^{\rm AGN}$ is the BLR luminosity in AGN frame} and $r_{\rm BLR}^{\rm AGN} = 0.1(L_{\rm BLR}^{\rm AGN}/10^{45}\rm erg \,  s^{-1})^{1/2}$~pc is the characteristic radius of the BLR in the AGN frame. The energy spectrum of the BLR radiation is taken to be that of black body with a peak at $\approx3.93k_{\rm B}T\Gamma/h\approx2\times10^{15}\Gamma$ Hz \citep{2008MNRAS.386..945T}. {The inner blob is assumed  to be located close to or inside the BLR, i.e.,  $r_{\rm in}^{\rm AGN}\sim r_{\rm BLR}^{\rm AGN}$.}

We neglect the radiation of protons that leave the inner blob, as they will cool adiabatically on a timescale $\sim r_{\rm BLR}^{\rm AGN}/(c\delta_{\rm D})\sim 10^6(r_{\rm BLR}^{\rm AGN}/10^{17}{\rm cm})(\delta_{\rm D}/20)^{-1}\,$s in the jet's comoving frame. This is generally much shorter than the photohadronic energy loss timescales (see Fig.~\ref{fig:timescale}), especially if one considers that the BLR radiation density, which dominates the proton photohadronic losses, decreases quickly at larger distances (see Eq.~\ref{eq:ublr}).

Using the steady-state proton distribution $N_{\rm p}(\gamma_{\rm p})$ and following \cite{2008PhRvD..78c4013K}, we compute the differential spectra of: $\gamma$-ray photons from neutral pion decays, electrons/positrons, and neutrinos produced in the photopion production and  BH pair-production processes. The spectrum of secondary electrons generated in the pair cascade is evaluated  using a semi-analytical method developed by \citet{2013ApJ...768...54B,2018ApJ...857...24W} \cite[see also][]{2015MNRAS.447.2810Y}. The synchrotron, SSC and EC emission of the pair cascade is also calculated.

Lastly, any very high energy (VHE) $\gamma$-ray photons that escape from the inner or the outer blobs will be absorbed by the extragalactic background light (EBL).  To account for this attenuation in the GeV-TeV band, we use the EBL model of \cite{2011MNRAS.410.2556D}.

\begin{table*}
\caption{Parameters for SED fitting with the two-zone $p\gamma$ model. {We assume a large distance of the outer blob to the SMBH (e.g., $\sim 10\,$pc$\gg r_{\rm BLR}^{\rm AGN}$) so that it is not influenced by the BLR radiation.}\label{tab}}
\centering
\begin{tabular}{@{}llllllllll@{}}
\hline\hline 
Free Parameters	&	$\delta_{\rm D}$	&	$B$	&	$R_{\rm out}$	&	$R_{\rm in}$	&	$L_{\rm e, inj}$	&	$r^{\rm AGN}_{\rm in}$ 	&    $n_{\rm e, 1}$	  & $n_{\rm e, 2}$	&	$\gamma_{\rm e, b}$	\\
 & &  [G] &  [cm] &  [cm] &  [erg/s] & [pc] & & \\
 \hline
Values	&	26.5	&	0.11	&	$4\times10^{16}$	&	$1\times10^{16}$	&	$2\times10^{42}$	&	0.05	&  1.4	& 4	&	$8.1\times10^{3}$	\\
\hline\hline
Fixed/Derived &  $L_{\rm BLR}^{\rm AGN}$ & $L_{\rm p, inj}$ & $\gamma_{\rm e, min}$ & $\gamma_{\rm e, max}$ & $n_{\rm p}$ & $\gamma_{\rm p, min}$ & $\gamma_{\rm p, max}$ & $L_{\rm e, in}^{\rm k}$ & $L_{\rm p,in}^{\rm k}$ \\
parameters & [erg/s] & [erg/s] & & & & & & [erg/s] & [erg/s] \\
 \hline
Values & $5\times10^{43}$ & $1.97\times10^{44}$ & 50 & $10^7$ & 2 & 1 & $7.04\times10^{6}$ & $3.83\times10^{43}$ & $1.3\times10^{47}$ \\
\hline
\end{tabular}
\tablecomments{The kinetic luminosity in relativistic electrons $L_{\rm e, in}^{\rm k}$ and in relativistic protons $L_{\rm p,in}^{\rm k}$ in the inner blob are calculated as $L_{\rm e,in}^{\rm k} \simeq \pi R_{\rm in}^2\delta_{\rm D}^2 m_{\rm e}c^3\int \gamma_{\rm e}N_{\rm e}(\gamma_{\rm e})d\gamma_{\rm e}$ and $L_{\rm p, in}^{\rm k} \simeq \pi R_{\rm in}^2\delta_{\rm D}^2 m_{\rm p}c^3\int \gamma_{\rm p}N_{\rm p}(\gamma_{\rm p})d\gamma_{\rm p}\simeq (3/4)\delta_{\rm D}^2L_{\rm p, inj}$. Due to the strong cooling of electrons, there is no simple relation between electrons' injection luminosity and its kinetic luminosity.}
\end{table*}

\subsection{Parameters constraints}
To reduce the number of free parameters as much as possible, we assume that the two regions have the same electron luminosity ($L_{\rm e, inj})$, the same proton luminosity ($L_{\rm p, inj})$, as well as the spectral shape of injected particles. They also have the same magnetic field $B$, and they move with the same Doppler factor $\delta_{\rm D}$. 
With such a setup\footnote{We discuss how a different choice of the model parameters for the two blobs would affect our results in Appendix~\ref{appendixA}.}, our two-zone $p\gamma$ model has sixteen free parameters:
six for the two radiation zones ($\delta_{\rm D}$, $B$, $R_{\rm out}$, $R_{\rm in}$, $r^{\rm AGN}_{\rm in}$ and $L_{\rm BLR}^{\rm AGN}$), six for the injected primary relativistic electrons in the two blobs ($L_{\rm e, inj}$, $n_{\rm e, 1}$, $n_{\rm e, 2}$, $\gamma_{\rm e, min}$, $\gamma_{\rm e, b}$ and $\gamma_{\rm e, max}$) and four for the injected relativistic protons in the inner blob ($L_{\rm p, inj}^{\rm AGN}$, $n_{\rm p}$, $\gamma_{\rm p, min}$ and $\gamma_{\rm p, max}$). Our strategy for reducing  further the number of free parameters is described below.  
\begin{enumerate}
\item We set $L_{\rm BLR}^{\rm AGN}=5\times 10^{43}\rm erg~s^{-1}$, as estimated by \cite{2019MNRAS.484L.104P}. The characteristic size of the BLR is then $r_{\rm BLR}^{\rm AGN}=0.02\,$pc. 

\item In leptonic models, the minimum electron Lorentz factor $\gamma_{\rm e, min}$ is usually constrained by the hard X-ray data,  since this is the energy band where the low-energy tail of the SSC emission emerges \citep{2008MNRAS.385..283C, 2016ApJ...831..142M}. However, when hadronic interactions are considered, the synchrotron radiation of secondary electrons generated in the pair cascade will also contribute to the hard X-ray band. Therefore, $\gamma_{\rm e, min}$ cannot be constrained  in a straightforward manner by the SED fitting. In our calculations, we therefore adopt a typical value of $\gamma_{\rm e, min}=50$. 

\item We set the maximum electron Lorentz factor $\gamma_{\rm e, max}=10^7$, since it does not have a major impact on our model results. 

\item
We set the power-law index of the proton spectrum at injection to be $n_{\rm p}=2$.

\item We adopt $\gamma_{\rm p, min}=1$ for the minimum proton Lorentz factor. Its exact value will not affect our fitting results, but only the energetic requirements.

\item Generally, $\gamma_{\rm p, max}$ can be obtained by equating the acceleration timescale to the minimum of the energy loss and escape timescales. The acceleration timescale can be evaluated by \citep{2004PASA...21....1P, 2007Ap&SS.309..119R}
\begin{equation}
t_{\rm acc}\simeq \frac{\alpha r_{\rm L}}{c} \simeq \frac{\alpha \gamma_{\rm p}m_{\rm p}c}{eB},
\end{equation}
where $r_{\rm L}$ is the Larmor radius of the relativistic proton {and $\alpha(>1)$ is a model parameter depending on the detailed mechanism. Since the detected neutrino energy is most likely $\lesssim$PeV, the maximum proton energy in the jet's comoving frame should be larger than $\sim 1(\delta_{\rm D}/20)^{-1}$\,PeV. Using $\alpha=50$ \citep{1983A&A...125..249L}}  and assuming that the acceleration is hindered by physical escape from the system, the maximum proton Lorentz factor is: 
\begin{eqnarray}
\gamma_{\rm p, max}&=&\frac{eBR_{\rm in}}{\alpha m_{\rm p}c^2}\nonumber \\ &  \simeq & 6.4\times10^7\left(\frac{\alpha}{50}\right)^{-1}
\left(\frac{B}{1\rm G}\right)\left(\frac{R_{\rm in}}{10^{16}\rm cm}\right)
\end{eqnarray}
which satisfies the condition to produce PeV neutrinos. 

\item Assuming that the SMBH mass of TXS 0506+056 is $10^9M_{\odot}$, the Eddington luminosity of the SMBH is $L_{\rm Edd}=1.3\times10^{47}\rm erg/s$. We set the kinetic proton luminosity for both blobs to be the Eddington luminosity, or an injection luminosity $L_{\rm p, inj}\simeq L_{\rm Edd}/\delta_{\rm D}^2$ in the blob. 
\end{enumerate}
Finally, the number of free parameters is reduced to nine, namely $\delta_{\rm D}$, $B$, $R_{\rm out}$, $R_{\rm in}$, $L_{\rm e, inj}$, $r^{\rm AGN}_{\rm in}$, $n_{\rm e, 1}$, $n_{\rm e, 2}$ and $\gamma_{\rm e, b}$.
\subsection{Application to TXS~0506+056 and IC-170922A}\label{sec:modeling}
\begin{figure}[htbp]
\centering
\includegraphics[width=1\columnwidth]{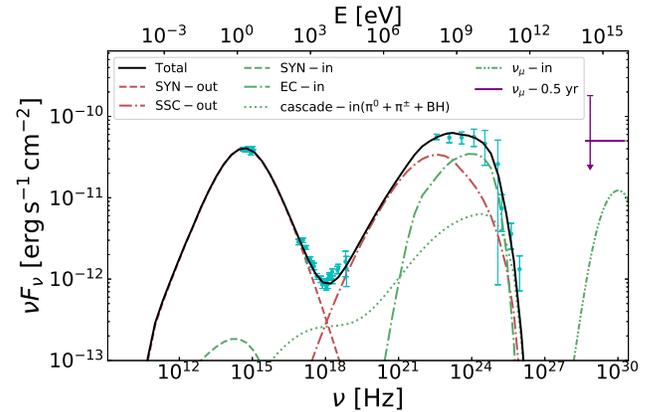}
\caption{The (quasi-)simultaneous SED of TXS~0506+056 within two weeks of the detection of IC-170922A (on MJD~58019). Data (colored symbols) are taken from \cite{2018Sci...361.1378I}. The dashed red and dotted-dashed red curves represent the synchrotron and SSC emission from primary electrons in the outer blob, respectively. The dotted green and  dot-dashed green curve represents the synchrotron emission and EC emission of primary electrons in the inner blob respectively, while the dotted green curve represents the emission from pair cascades in the inner blob. The double dotted-dashed green curve shows the muon neutrino energy spectrum. {The SSC emission of primary electrons in the inner blob is too low to be visible in this figure.} The solid black curve is the total emission from the outer blob and II. {The representative solid purple lines show the required $\nu_\mu+\bar{\nu}_\mu$ neutrino flux to produce one EHE alert event like IC-170922A over a period of half year, assuming a neutrino spectrum of $E^{-2}$}.\label{fig:sed}}
\end{figure}

We apply the two-zone $p\gamma$ model  to the 2017 multi-messenger observations of TXS~0506+056.   
 Our model-predicted SED and neutrino spectrum for the parameters shown in Table~\ref{tab} are presented in Fig.~\ref{fig:sed}.   
The low-energy bump in the SED is  explained  by the synchrotron emission of primary electrons in the outer blob, while the hard X-ray flux and part of the GeV flux is attributed to their SSC emission. The primary electrons in the inner blob cool very fast in the Doppler-boosted BLR radiation field. The synchrotron radiation of these electrons is suppressed, as they radiate away their energy mainly through the EC process\footnote{The energy density of BLR at $r^{\rm AGN}_{\rm in}$ is $\sim 1.6\,  \rm erg~cm^{-3}$, which is much higher than the magnetic energy density, i.e. $\sim 0.0004 \, \rm erg~cm^{-3}$.} (see also  Fig.~\ref{fig:timescale}), which makes an important contribution to the {\it Fermi}-LAT $\gamma$-ray flux. In addition, the emission from the pair cascade mainly accounts for the TeV emission.  We note that this is not a unique representation of the SED, as different parameter sets can lead to equally good descriptions of the blazar spectrum. We refer readers to Appendix~\ref{appendixA} for a detailed discussion on this topic.

By convolving the neutrino flux shown in Fig.~\ref{fig:sed} with the effective area of IceCube's EHE event alert system \citep{2018Sci...361.1378I}, we derive the predicted neutrino detection rate. This is $\sim 0.3\, \rm yr^{-1}$ in the range of $0.2-7.5$\,PeV, which is one order of magnitude higher than the one predicted by standard one-zone models \citep{2018ApJ...864...84K, 2019MNRAS.483L..12C, 2019NatAs...3...88G}. 

To compute the number of predicted muon neutrinos from TXS 0506+056, one needs to know the duration of neutrino production in addition to the predicted production rate.
Since only one neutrino event (IC-170922A) was detected during the $\sim$six-month long  $\gamma$-ray flare of TXS~0506+056 in 2017, the neutrino emission period cannot be securely determined from the observations. From the theoretical point of view,
in the two-zone $p\gamma$ model considered here, the period of $\gamma$-ray flaring activity and enhanced neutrino production can, in principle, be different. The reason is that the $\gamma$-ray emission may be explained solely by the radiative processes of primary electrons in the outer blob, while the neutrino emission is mainly produced in the inner blob via $p\gamma$ interactions of relativistic protons with the BLR photons. If the inner blob is moving relativistically towards the observer, the neutrino emission at the predicted rate of $\sim 0.3$~yr$^{-1}$ would only last for a time equal to the maximum of the following two characteristic timescales: the light-crossing timescale of the inner blob (i.e., $\sim R_{\rm in}/\delta_{\rm D}c$) and the time needed for the inner blob to travel beyond the BLR (as seen in the observer's frame), i.e., a few times $r_{\rm in}^{\rm AGN}/\delta_{\rm D}^2c$, which are both less than one day. This short duration would result in a very low detection probability in spite of the high neutrino production rate.  
In order to obtain higher detection probabilities within our model, the neutrino production rate of $\sim 0.3$~yr$^{-1}$ should be sustained for several months. This might be realized if multiple blobs, like the inner blob, would form close to or inside the BLR as a result of continuous dissipation for a  period of several months. Alternatively, the inner blob could be associated with dissipation happening at a quasi--stationary feature (such as re-collimation shock, e.g., \citealt{2011A&A...531A..95F}) close to or inside the BLR, lasting the same amount of time.
 
\section{Comparison with one-zone models}\label{compare}
The main difference between the two-zone $p\gamma$ model from the standard one-zone models is that the emission accounting for the low-energy bump of the blazar SED originates from a different region than the one where neutrinos are produced. This is the key to overcome the constraint from the X-ray flux on the cascade emission and hence increase the maximal predicted neutrino flux. To further explain this point, let us first review briefly the reason why one-zone $p\gamma$ models always result in lower neutrino fluxes. 

\subsection{One-Zone SSC Model}
For a typical Doppler factor of the blazar jet $\delta_{\rm D}\sim 10$, PeV neutrinos are produced through the $p\gamma$ process most efficiently on soft X-ray photons (as viewed in the jet's comoving frame). In the one-zone SSC model, the target photon field is the synchrotron radiation of primary electrons, which accounts for the low-energy bump in the SED of TXS~0506+056. The
comoving synchrotron spectrum peaks in the infrared, thus resulting in a very low number density in the X-ray band. Consequently, the PeV neutrino production efficiency is low, as it has also been pointed out in previous studies. There are in principle two potential ways of overcoming this problem and of increasing the neutrino flux. 

The first one is to increase the $p\gamma$ interaction efficiency $f_{p\gamma}$. This can be achieved by increasing the number density of the synchrotron radiation in the comoving frame by employing a relatively low Doppler factor $\delta_{\rm D}$ and/or small blob size $R$. At the same time, however, the opacity for $\gamma\gamma$ absorption of high-energy gamma rays by the synchrotron photons will also increase by the same factor (i.e., $\tau_{\gamma\gamma}\propto \delta_{\rm D}^{-4}R^{-1}$, \citealt{2012ApJ...755..147D, 2017MNRAS.470.1881P, 2019ApJ...871...81X}). Note that the synchrotron radiation provides abundant infrared photons which efficiently absorb co-produced gamma-ray photons with sub-PeV/PeV energy and deposit their energy via the pair cascade into the X-ray band \citep{2015MNRAS.447...36P,2016PhRvL.116g1101M}. 
Although the neutrino luminosity is $L_\nu\propto f_{p\gamma}L_{\rm p}$, the X-ray luminosity from the cascade is $L_{\rm X}\propto \tau_{\gamma\gamma}f_{p\gamma}L_{\rm p}\propto f_{p\gamma}^2L_{\rm p}$. Thus, if one attempts to increase $f_{p\gamma}$ to produce a higher neutrino flux, the  X-ray flux of the EM cascade will unavoidably increase as $f^2_{p\gamma}$. The \textit{Swift}/XRT and \textit{NuSTAR} X-ray observations of TXS~0506+056 are placing strong constraints on $f_{p\gamma}$ \citep{2018ApJ...864...84K, 2018ApJ...865..124M,2019NatAs...3...88G}.

Alternatively, one can employ a high proton luminosity to counterbalance the low neutrino production efficiency.
Given that $L_{\rm X}\propto f_{p\gamma}^2L_{\rm p}$, a high neutrino flux can be obtained simultaneously with a low X-ray flux from cascade by invoking a sufficiently small $f_{p\gamma}$ and a sufficiently large $L_{\rm p}$. However, the required proton luminosity would be highly super-Eddington \citep[see also][]{2015MNRAS.448.2412P}. 

Practically speaking, the fitting of the high-energy bump of the SED with the SSC emission of primary electrons requires a proper energy density of the synchrotron radiation \citep[e.g.,][]{1997A&A...320...19M}. Thus, the neutrino production efficiency is more or less fixed.
The neutrino detection rate predicted by previous studies of the 2017 multi-messenger flare of TXS~0505+056 within the one-zone SSC model, when the pair cascade emission saturates the X-ray data, is $\sim 0.01-0.03\rm \, yr^{-1}$. Still,
the required proton luminosity  exceeds the Eddington limit by several tens to a few hundred times (assuming a black hole mass of $M_{\rm BH}=10^9M_\odot$).

\subsection{One-Zone EC model}
In the one-zone EC model, external photons such as photons from the BLR \citep[e.g.,][]{2001PhRvL..87v1102A, 2014JHEAp...3...29D, 2014PhRvD..90b3007M}, the accretion disk or accretion flow \citep[e.g.,][]{1991PhRvL..66.2697S, 1999MNRAS.302..373B, 2004PhRvD..70l3001A, 2015ApJ...806..159K, 2019MNRAS.483L.127R}, or the sheath region of the jet \citep[e.g.,][]{2014ApJ...793L..18T, 2015MNRAS.451.1502T}, can significantly enhance the $p\gamma$ interaction efficiency and the neutrino flux, and thus may reduce the required  proton luminosity to sub-Eddington levels. This can be understood as follows. First, the external photon energy, which typically lies in the optical/UV band, in the blob comoving frame appears (due to Doppler boosting) in the soft X-ray band, which is favorable for the production of sub-PeV/PeV neutrinos. Second, the photon number density in the jet's comoving frame is enhanced, thus leading to higher neutrino production efficiency. However,  given the high density of the Doppler-boosted external photon field, {a strong magnetic field of $\sim 2 \, (u_{\rm BLR}/{\rm erg \, cm^{-3}})^{1/2}\,$G} is required to channel sufficient electron energy into synchrotron radiation and explain the low-energy bump of the SED. The produced synchrotron photons can still result in a high opacity for high-energy gamma rays (see Fig.~\ref{fig:tau}), and then efficiently transfer energies of EM secondaries that are co-produced with neutrinos to the pair cascade. At the same time, the strong magnetic field enhances the synchrotron emission of the pair cascade in the X-ray band. 
Of course, one can adjust the model parameters to maximize the neutrino flux by reducing the magnetic field and the density of external photon field simultaneously. Nevertheless, on the premise of fitting the SED, a low neutrino detection rate similar to that in the one-zone SSC model is always obtained \citep{2018ApJ...864...84K}.

\subsection{Two-Zone $p\gamma$ model}
In the two-zone $p\gamma$ model, the low-energy bump of the SED is produced by primary electrons in the outer blob, where the influence of the BLR radiation is negligible. Unlike the one-zone EC model, one does not have to adopt a strong magnetic field for the inner blob to explain the low-energy bump of the SED, as the latter is accounted for by the synchrotron radiation of primaries in the outer blob. The synchrotron radiation of primary electrons in the inner blob can be therefore suppressed by the EC process on the BLR radiation field, so that the $\gamma\gamma$ absorption for sub-PeV/PeV gamma rays is low. A large fraction of the energy injected into very high energy secondaries produced by the $p\gamma$ interactions will not be deposited in the EM cascade. 
Furthermore, secondary electrons generated in the cascades radiate mainly through the EC process. Hence, the X-ray flux from cascades is much lower than that in one-zone models, and subsequently allows for a much higher neutrino flux. Note that, the radiation field in the outer blob has little contribution to the absorption of gamma-rays escaping the inner blob (see Appendix~\ref{appendix:opa} for details).  

In Fig.~\ref{fig:tau}, we compare the opacity of the $\gamma\gamma$ absorption in the inner blob as a function of the photon energy in the observer's frame for the two-zone $p\gamma$ model (with parameters given in Table~\ref{tab}), the one-zone SSC model with parameters from \citet{2019MNRAS.483L..12C}\footnote{i.e., $B = 0.2 \, {\rm G}, R=1.5\times10^{16} {\rm cm}, \delta_{\rm D}=34, L_{\rm e, inj}=4\times10^{42}\, {\rm erg \, s^{-1}}, n_{\rm e, 1}=2, \gamma_{\rm e, min}=500, \gamma_{\rm e, max}=1.5\times10^4$.}, and the one-zone EC model with parameters from \citet{2018ApJ...864...84K}\footnote{ i.e., $ B = 0.4 \, {\rm G}, R=10^{17}{\rm cm}, \delta_{\rm D}=24.2, L_{\rm e, inj}=2.2\times10^{42} \, {\rm erg \, s^{-1}}, n_{\rm e, 1}=1.9, n_{\rm e, 2}=3.6, \gamma_{\rm e, min}=1, \gamma_{\rm e, b}=5\times10^3, \gamma_{\rm e, max}=8\times10^4, u_{\rm ext}=0.033 \, \rm erg \, cm^{-3}$.}. The inner blob is optically thin to the attenuation of PeV photons in the two-zone model, but is predicted to be optically thick in the other two models, as shown in Fig.~\ref{fig:tau}.


\begin{figure}[htbp]
\centering
\includegraphics[width=1\columnwidth]{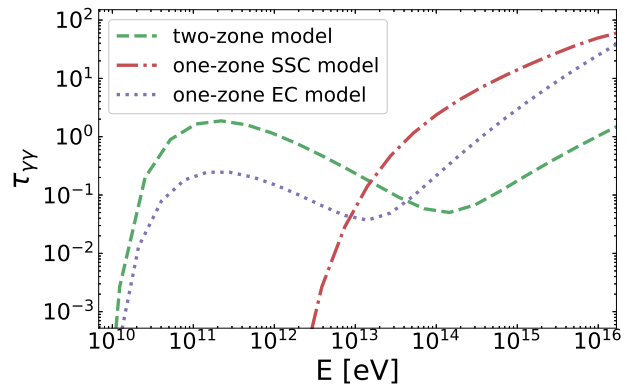}
\caption{Comparison of the opacity of $\gamma\gamma$ annihilation as a function of the photon energy in the observer's frame in the two-zone $p\gamma$ model (dashed green curve), one zone EC model (dotted purple curve) and one-zone SSC model (dash-dotted red curve).  See Section~\ref{compare} for detailed discussion.\label{fig:tau}}
\end{figure}
\section{Discussion and conclusions}\label{dis}
\subsection{Influence of the dusty torus}\label{dt}
According to the unification scheme of AGN, a dusty torus (DT) is also expected to be present at large distances from the SMBH (i.e., $\sim1$~pc).
Its radiation is treated as an isotropic blackbody spectrum with a peak at $3\times10^{13}\Gamma$ Hz \citep{2007ApJ...660..117C} in the jet comoving frame. Denoting the DT luminosity by $L_{\rm DT}^{\rm AGN}$ in the AGN frame, the energy density of the DT radiation in the jet comoving frame varies with the distance to the SMBH ($r^{\rm AGN}_{\rm in}$), and it can be written as \citep{2012ApJ...754..114H}:
\begin{equation}
u_{\rm DT} \approx \frac{\Gamma^2L_{\rm DT}^{\rm AGN}}{4\pi (r_{\rm DT}^{\rm AGN})^2c[1+(r^{\rm AGN}_{\rm in}/r_{\rm DT}^{\rm AGN})^3]},
\end{equation}
where $r_{\rm DT}^{\rm AGN} = 2.5(L_{\rm DT}^{\rm AGN}/10^{45}\rm erg \, s^{-1})^{1/2}$pc is the characteristic radius of the DT in the AGN frame.  This results in $r_{\rm DT}\simeq 0.56\,$pc assuming that $L_{\rm DT}^{\rm AGN}=L_{\rm BLR}^{\rm AGN}$ and the outer blob is not expected to be influenced by the DT radiation.


\begin{figure}[htbp]
\centering
\includegraphics[width=1\columnwidth]{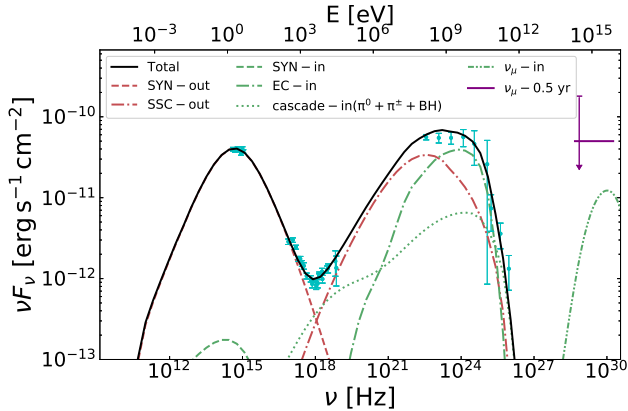}
\caption{Same as in Fig.~\ref{fig:sed}, except taking into account the presence of a dusty torus with luminosity $5\times 10^{43}\rm erg/s$. All other parameters are identical to those shown in Tab.~\ref{tab}.} \label{fig:ir}
\end{figure}

The typical energy of the DT radiation (as viewed in the jet's comoving frame) falls in optical band, thus acting as a target photon field for BH pair production and $\gamma\gamma$ pair production. As a result, inclusion of the DT in the radiative transfer calculations for the inner blob, leads to higher X-ray fluxes due to the injection of more 
sub-TeV/TeV electrons into the pair cascade.

On the other hand, the DT radiation field will suppress the synchrotron radiation of the sub-TeV/TeV electrons, which would produce X-ray emission. These two effects influence the X-ray emission from pair cascades in opposite directions and somewhat counteract each other. Indeed, from  Fig.~\ref{fig:ir}, we see that the maximal neutrino flux  obtained (i.e., when the observed X-ray flux is saturated by the cascade emission)  with inclusion of the DT radiation is comparable to that obtained previously  in the absence of the DT (see e.g. Fig.~\ref{fig:sed}). Therefore, whether a DT exists in this source or not has a minor influence on the SED fitting and neutrino flux. 

\subsection{Possible application to the 2014/2015 neutrino flare}
\label{subsec:archival_flare} 
The historical IceCube data in the position of TXS~0506+056 show a $3.5\sigma$ excess of high-energy muon neutrinos between September 2014 and March 2015, with $13\pm 5$ events in the range of $32\,\rm TeV - 3.6\,PeV$ \citep{2018Sci...361..147I}. This corresponds to an all-flavor isotropic-equivalent neutrino luminosity of $1.2_{-0.4}^{+0.6}\times 10^{47}\rm erg \, s^{-1}$. During this period, TXS~0506+056 was not exhibiting any flaring activity in electromagnetic radiation (from radio to gamma-rays), thus suggesting that the neutrino excess could not have been produced by an
enhancement of the intensity of target radiation fields for the $p\gamma$ process. Thus, we can rule out the conventional one-zone model for the enhanced neutrino emission \citep{2019ApJ...874L..29R, 2018arXiv180900601W}. Particularly, the isotropic-equivalent gamma-ray luminosity above 0.1\,GeV is only one fifth of the neutrino luminosity. Such a feature has been interpreted as strong internal absorption of gamma-ray photons by X-ray radiation \citep{2018arXiv180900601W, 2018arXiv181205654R}, which might arise from the thermalized radiation inside a massive cloud that enters the jet in the framework of two-zone $pp$ model \citep{2018arXiv180900601W}. Here, we briefly discuss the potential of the two-zone $p\gamma$ model for explaining the 2014/2015 neutrino excess. 

There are two basic observational constraints: (1) the blazar is not exhibiting any EM flare during the neutrino excess, so there is no good physical reason for the proton injection luminosity  to be super-Eddington; (2) unless one an extremely large proton-to-electron ratio (i.e., baryon loading factor) for particle injection, the radiation of injected electrons 
needs to be suppressed or absorbed to avoid producing a an EM flare. One possibility is to consider that the inner blob forms close to a hypothetical hot X-ray corona at a distance comparable to or a few times larger than the gravitational radius of the SMBH (i.e., $r_{\rm co}\sim 10^{14}-10^{15}\,$cm). The corona's radiation is thought to be produced by Compton upscattering of UV and optical photons from the SMBH accretion disk by hot electrons. It typically has a power-law spectrum with photon index $\alpha \simeq 1-2$ and a cutoff in the range of $\lesssim 100$\,keV for AGN with high Eddington ratio~\citep[e.g.,][though for non-blazar AGN]{2018ApJ...866..124K,2018MNRAS.480.1819R}. For simplicity, we neglect the effects of general relativity from the following qualitative discussion. 
Let us assume the corona's luminosity to be $L_{\rm X}=10^{43}{\rm erg \, s^{-1}}$ at 1\,keV which is comparable to that of the BLR. By using the $\Delta$ resonance approximation for the $p\gamma$ cross section,
we find the production efficiency of neutrinos to be $f_{\nu}
\simeq (3/8)\xi_\Delta\sigma_\Delta n_{\rm X} r_{\rm co}\sim 5\times 10^{-3}(L_{\rm X,1keV}/10^{43}{\rm erg~s^{-1}})(r_{\rm co}/10^{14}~{\rm cm})^{-1}(E_\nu/15~{\rm TeV})^{\alpha-1}$, where $n_{\rm X}=L_{\rm X}/4\pi r_{\rm co}^2c \epsilon_{\rm X}$ is the X-ray photon number density, $\xi_\Delta=0.2$ is the inelastic coefficient of the $p\gamma$ interactions, the prefactor $3/8$ is the fraction of the energy lost by the proton that goes into neutrinos, and $\sigma_\Delta\simeq 4\times 10^{-28}\rm cm^{-2}$ is the cross section of the $\Delta$ resonance. Assuming protons are injected with a differential kinetic luminosity $E_{\rm  p}^2L_{\rm p}^{\rm k}(E_{\rm p})=5\times 10^{45}\rm erg \,s^{-1}$ (so that the proton's kinetic luminosity integrated from 1\,GeV to 100\,PeV is still sub-Eddington), we obtain the all-flavor differential neutrino luminosity in the AGN frame as $E_\nu^2L_\nu(E_\nu)=f_\nu(E_\nu)E_{\rm p}^2L_{\rm p}^{\rm k}(E_{\rm p}=20E_\nu)=3\times 10^{43}(E_\nu/15{\rm TeV})^{\alpha-1} \rm erg \, s^{-1}$. Given a typical jet's Doppler factor $\delta_{\rm D}=20$ and the corona's spectral index $\alpha=1$, the isotropic-equivalent neutrino luminosity in the range of $32\,{\rm TeV}-3.6\,{\rm PeV}$ can be found by multiplying a factor of $\sim \delta_{\rm D}^2\ln(3600/32)$, resulting in $L_{\nu, \rm iso}\sim 6\times 10^{46}\rm erg/s$, which is consistent with the observation. 

The synchrotron radiation of injected electrons are suppressed by the EC process on the X-ray radiation as long as the magnetic field in jet's comoving frame is  $\ll 10^4\,$G (i.e., $U_{\rm B}\ll U_{\rm co}$). The IC radiation of electrons would produce GeV emission but the $\gamma\gamma$ opacity in the X-ray radiation field is $\tau_{\gamma\gamma}=\sigma_{\gamma\gamma}L_{\rm X}/4\pi r_{\rm co}c\epsilon_{\rm X, 1keV}\simeq 16 (L_{\rm X, 1keV}/10^{43}\rm erg~s^{-1})(r_{\rm co}/10^{14}{\rm cm})^{-1}$. Such a large gamma-ray opacity is consistent with the fact that there is no associating gamma-ray flare with the neutrino flare. It is therefore possible to explain the EM and neutrino data of the 2014/2015 period within our model.

The absorbed GeV flux would be reprocessed into MeV band through the IC scattering on the external X-ray radiation field \citep{2018arXiv180900601W,2019ApJ...880...40I, 2019arXiv190404226M}. During the 2014/2015 neutrino excess, there is no observation at MeV band that can constrain the model. Future MeV--GeV detectors such as e-ASTROGAM \citep{2018JHEAp..19....1D},  AMEGO \citep{2017ICRC...35..798M} and PANGU \citep{2016SPIE.9905E..6EW} would be able to test the model in a similar event. 

\subsection{The diffuse neutrino emission in the two-zone $p\gamma$ model}
As we have shown in Section~\ref{sec:modeling}, the two-zone model can enhance the expected neutrino flux from TXS~0506+056. One may wonder how the predicted diffuse neutrino flux compares to the IceCube observations \citep{2017ApJ...835...45A}, if our model is generalized to the blazar population.

We here provide a rough estimate for the contribution of the blazar population to the diffuse neutrino flux in the context of the two-zone $p\gamma$ model. To do so, we simply use a scaling relation between the blazar gamma-ray luminosity and the expected neutrino luminosity,
based on the results obtained in this work for the flaring state of TXS~0506+056. For simplicity, we do not exclude from our analysis the true BL Lac objects, which have no or weak BLR radiation, although their neutrino emission would also be weak. 
Therefore, the  estimation of the diffuse neutrino flux in the two-zone model should be regarded as a crude upper limit.
 
According to the population study of GeV blazars by \citet{2014ApJ...780...73A}, the gamma-ray energy production rate of the entire population peaks at redshift $z\sim 1-2$ with a value of $\dot{W}_{\gamma}=10^{46}\rm \,erg~ Mpc^{-3}yr^{-1}$. For TXS~0506+056, the isotropic-equivalent gamma-ray flare luminosity is $\sim 10^{47}\rm \,erg~ s^{-1}$, and the muon neutrino luminosity at the spectral peak (i.e., PeV) predicted in the two-zone $p\gamma$ model is about $3\times 10^{45}\rm \,erg~s^{-1}$. If we scale the PeV muon neutrino production rate with the gamma-ray production rate following the case of TXS~0506+056, we obtain a PeV muon neutrino production rate of $\dot{W}_{\nu}({\rm PeV})=1.5\times 10^{43}(f_{\rm fl}/0.05)\rm \,erg~Mpc^{-3}yr^{-1}$. Here $f_{\rm fl}$ is the duty factor of blazar flares \citep{2018ApJ...865..124M}, which is introduced by considering that the neutrino emission of blazar is dominated by the flare state. It then leads to a diffuse neutrino flux at PeV of
\begin{equation}
\begin{split}
\Phi_\nu({1\,\rm PeV})& \approx \frac{cf_{
\rm fl}\dot{W}_{\nu}({1\,\rm PeV})}{4\pi H_0} \\
& =10^{-8}\left(\frac{f_{\rm fl}}{0.05}\right)\rm GeV~cm^{-2}s^{-1}sr^{-1}.
\end{split}
\end{equation}
 Note that the differential 90\% C.L. upper limit of diffuse muon neutrino flux from blazars is $10^{-8}\rm \,GeVcm^{-2}s^{-1}sr^{-1}$ at PeV  \citep{2017ApJ...835...45A}. Therefore, our model is consistent with the observational constraints as long as the neutrino flare duty factor $f_{\rm fl}$ is $\lesssim 0.05$. Alternatively, even if the relevant flare duty factor is larger, the two-zone leptonic model is consistent with the observed upper limit of blazars to the IceCube neutrino flux, as long as only a fraction of gamma-ray flares in blazars are due to dissipation (i.e., particle acceleration) close to or inside the BLR where the neutrino production efficiency is high. Indeed, the BLR usually extends out to a distance of $\sim 0.1-1\,$pc from the SMBH, and  only a small fraction of the jet is close to or inside the BLR given a jet length of $10-100\,$pc. If we assume dissipation or particle acceleration occur { with an equal probability per unit length} along the jet, the probability of forming a radiating zone inside the BLR should be $\sim 1-10\%$. { Our argument may be supported by analysis of \citet{2018MNRAS.477.4749C} who found that only 1 out of 10 FSRQs shows a feature of substantial attenuation in their gamma-ray spectrum, implying that only 10\% of the emitting regions of FSRQs are close to or inside their BLRs.}

\subsection{Distinguishing between the two-zone $p\gamma$ model and other models}
\citet{2019PhRvD..99f3008L} studied a two-zone $pp$ model in which neutrinos are produced through $pp$ collisions between accelerated protons and BLR clouds that enter the jet. This scenario is very similar to our two-zone $p\gamma$ model in terms of the physical separation of the regions responsible for the neutrino  and low-energy photon emissions.
Additionally, in the two-zone $pp$ model, the BLR clouds that enter the jet will be fully ionized and, subsequently, the generated free electrons may further reduce the X-ray flux by deflecting X-ray photons into other direction from our line of sight via Compton scattering. As a result, the two-zone $pp$ model is less constrained by the X-ray observations and can, in principle, lead to even higher neutrino flux than the one predicted in the two-zone $p\gamma$ model.

Both the $pp$ and $p\gamma$ two-zone models can reproduce the SED of the blazar and generate a comparatively high neutrino flux. Yet, the predicted TeV neutrino spectrum can be used as a model diagnostic.
The neutrino spectrum generated by $pp$ collisions basically follows the spectrum of the parent protons, which is generally believed to be a power law extending down to GeV energies. On the contrary, the neutrino spectrum generated by $p\gamma$ interactions peaks around PeV energies and the flux drops quickly at lower energies because the number density of target photons needed to produce these low-energy neutrinos is too low. Thus, if the neutrino spectrum can be measured or constrained in the TeV--PeV range in a blazar flare similar to that associated with IC-170922A in the future, we can distinguish between these two mechanisms. To achieve this goal, next-generation neutrino telescopes, such as IceCube-Gen2 with several times larger effective area compared to IceCube, are required.

Although the  neutrino flux predicted by one-zone models for the 2017 flare of TXS~0506+056 is low, the model predictions are still consistent with the detection of one event. Despite the higher neutrino fluxes predicted in the framework of two-zone models, the expected number of muon neutrinos  during the 2017 gamma-ray flare is still $\lesssim 1$. Thus, we cannot currently distinguish one-zone models from two-zone models for  IC-170922A.
However, if the same blazar flare was to occur in the era of IceCube-Gen2, the expected number of muon neutrinos in two-zone models would easily exceed unity, in contrast to one-zone models.


\section{Conclusion}\label{con}
The neutrino event IC-170922A coincident with gamma-ray flare of blazar TXS~0506+056 has been explained as an upward fluctuation or as the first neutrino to be detected from an ensemble of faint neutrino emitting sources in one-zone models. In these models, the neutrino detection rate is strongly constrained by the X-ray observations, because X-ray emission is unavoidably produced by the synchrotron radiation of the secondary electrons developed in pair cascades accompanying the neutrino production. 

In an attempt to obtain a higher neutrino detection rate than the one predicted with one-zone models, we proposed a two-zone $p\gamma$ model for the 2017 blazar flare. 
According to our scenario, two physically distinct emitting regions (blobs) are present in the blazar jet. In the inner blob, which is located inside or close to the BLR, 
relativistic electrons mainly radiate/cool via the EC process and neutrinos are efficiently produced via $p\gamma$ interactions of relativistic protons on Doppler-boosted BLR photons. In the outer blob, which is located well beyond the BLR, 
relativistic electrons radiate mainly via synchrotron and SSC processes; the neutrino production efficiency in this emitting region is low.  We demonstrated that
the combined emission from these two radiation zones can fit the SED of the blazar flare satisfactorily. We showed that the predicted PeV $\nu_\mu+\bar{\nu}_\mu$ flux  during the flare period can be as high as $10^{-11}\rm \,erg$ $\rm cm^{-2}$ $\rm s^{-1}$, without invoking a super-Eddington luminosity for the  injected protons. 

Using the effective area of the IceCube EHE alert system, the predicted neutrino flux translates to a detection rate of $\sim 0.3\rm \,yr^{-1}$. The latter is about one order of magnitude larger than the one derived in one-zone models.  The reason behind this difference is that the two-zone model relaxes the constraints imposed on the neutrino flux in two ways:
\begin{enumerate}
\item The emitting zone responsible for the low-energy hump of the SED is physically separated from the neutrino emitting region. As a result, the $\gamma\gamma$ absorption opacity for high-energy photons is significantly reduced in the neutrino emitting zone. This suppresses the emission from the EM cascade by reducing the number of high-energy photons/electrons that participate in the pair cascade.

\item In the presence of a dense external radiation field, the X-ray emission, which arises from the synchrotron radiation of electrons in the pair cascade, is suppressed by the EC process. As a result, the radiation of secondary pairs is channeled preferentially to higher energies  than in X-rays.
\end{enumerate}
Currently, it is difficult to discriminate the two-zone $p\gamma$ model presented here from other scenarios proposed for the multi-messenger flare of TXS 0506+056. This may become possible with future high-energy neutrino detections from TXS~0506+056 or other blazars by next-generation neutrino telescopes with larger effective areas, such as IceCube-Gen2.

\acknowledgments

This work is supported by the National Key R$\&$D program of China under the grant 2018YFA0404200, and the NSFC under grants 11625312 and 11851304.  F.O. is supported by the Deutsche Forschungsgemeinschaft through grant SFB\,1258 ``Neutrinos and Dark Matter in Astro- and Particle Physics''. M.P. acknowledges support from the Lyman Jr.~Spitzer Postdoctoral Fellowship and NASA  Fermi 80NSSC18K1745. K.W. is supported by the China Postdoc Science Grant (No. 2019M650311).

\appendix
\section{Explaining the SED with other set of parameters}\label{appendixA}
In our default SED fit shown in Fig.~\ref{fig:sed},
we assumed the same magnetic field strength and injection electron spectrum for both blobs,  despite their large separation, 
with the only requirement that $R_{\rm out}\gtrsim R_{\rm in}$.
Our assumption implies that either the jet is not conical or that the size of the outer blob is much smaller than the jet's transverse radius at that location. The latter scenario has been also proposed to explain the fast variability of the optical blazar emission, if this is emitted at a large distance from the SMBH \citep{2011A&A...534A..86T}. 

Our default SED fit shown in Fig.~\ref{fig:sed} is not unique.
In fact, because of the multi-parameter nature of  two-zone models, it is often not possible to constrain their free parameters by SED fitting alone.
For the purposes of this work, it is sufficient to find a set of reasonable parameters for fitting the SED and demonstrate the feasibility of the model. Nevertheless, it is useful to discuss different parameter combinations that can still describe well the SED and, at the same time, correspond to different physical realizations. In the following, we consider two alternative scenarios.



\begin{figure}[htbp]
\centering
\includegraphics[width=0.45\columnwidth]{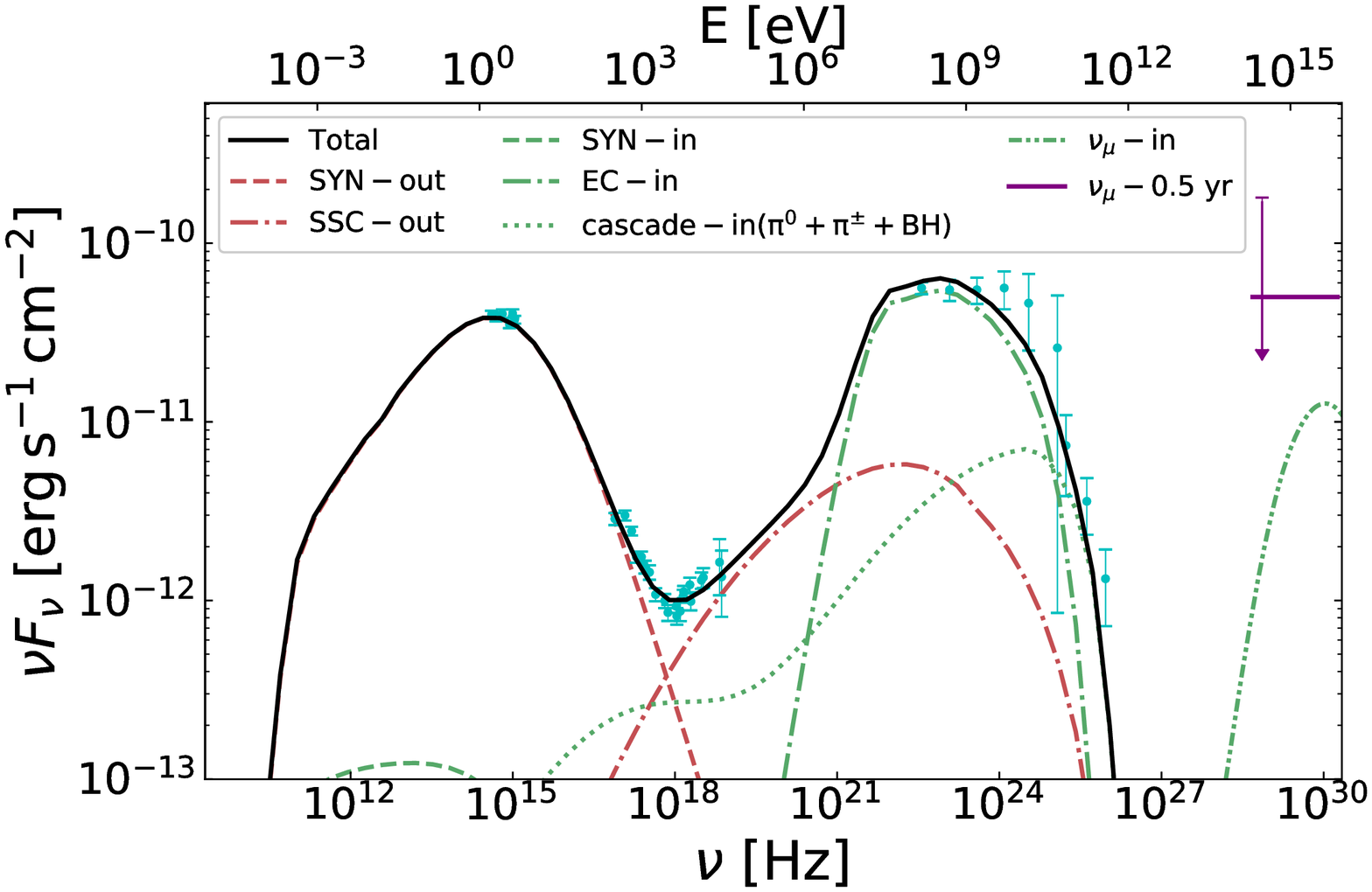}
\includegraphics[width=0.45\columnwidth]{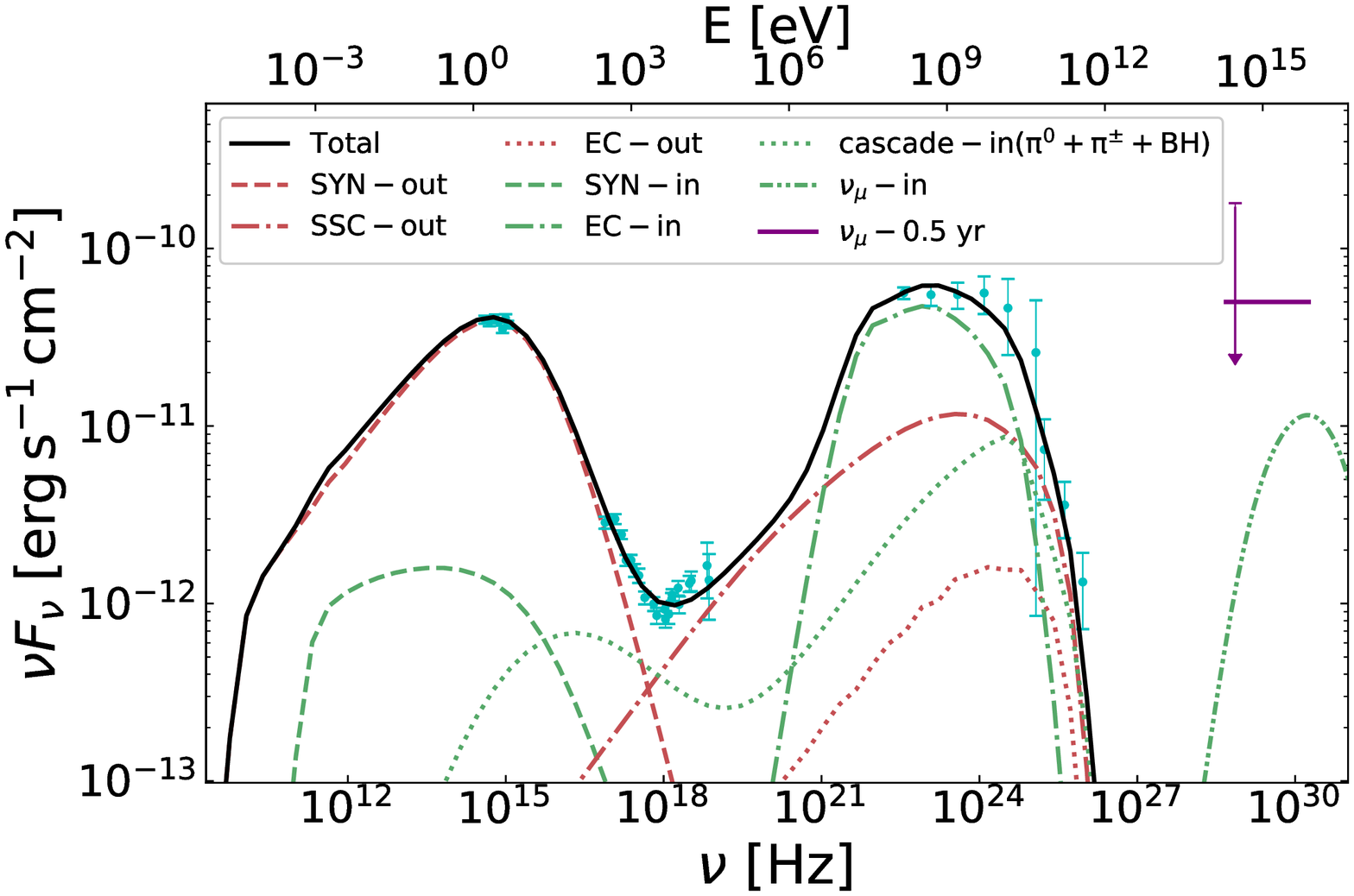}
\caption{Examples of SED fits obtained within the context of our two-zone model for different parameter sets. 
In both cases, the GeV gamma-ray flux is dominated by the EC emission from the inner blob, unlike the default SED fit shown in Fig.~\ref{fig:sed}. In the right panel, the magnetic field and the blob radius are scaled to the distance of the blob from the SMBH. The line styles have the same meaning   as in Fig.~\ref{fig:sed}. The parameter values used for the models shown in the left and right panels are summarized in Tables~\ref{tab2} and \ref{tab3}, respectively. \label{fig:alt_fit}}
\end{figure}

Let us first look into a case where the GeV flux is dominated by the EC emission from the inner blob. To achieve this, we simply suppress the SSC emission of the outer blob by adopting a few times larger radius  than the one used in our default fit (see Tables~\ref{tab} and \ref{tab2}). 
For the adopted parameters, the SSC flux from the outer blob is comparable to the gamma-ray flux in the  \textit{Fermi}-LAT band  at a low state. If there is no dissipation happening close to the BLR (i.e., without the inner blob), the outer blob can solely explain the SED in the low state, without expecting significant neutrino emission. Note that electrons cool quite fast in the inner blob (under the parameters in Table~\ref{tab2}, a cooling break will appear in the electron spectrum at $\gamma_{\rm cool}\approx75$), so the spectral index of the steady-state electrons responsible for GeV emission is $n_{\rm e,1}+1$ and thus we assume $n_{\rm e, 1}=2$ to account for the flat spectrum observed at GeV. The results are shown in the left panel of Fig.~\ref{fig:alt_fit}.

\begin{table*}
\caption{Model parameters for SED fit shown in the left panel of Fig.~\ref{fig:alt_fit}\label{tab2}}
\centering
\begin{tabular}{@{}lllllllllll@{}}
\hline\hline
Free parameters&$\delta_{\rm D}$&$B$&$R_{\rm out}$&$R_{\rm in}$&$L_{\rm e, inj}$&   $r^{\rm AGN}_{\rm out}$& $r^{\rm AGN}_{\rm in}$ &    $n_{\rm e, 1}$ & $n_{\rm e, 2}$&$\gamma_{\rm e, b}$\\
 & &  [G] &  [cm] &  [cm] &  [erg/s] & [pc] & [pc] & & & \\
 \hline
Values&26.5&0.12&$8\times10^{16}$&$2\times10^{16}$&$1.7\times10^{42}$  &10  &0.07&  2& 4&$1\times10^{4}$\\
\hline
\hline
Fixed/Derived  &  $L_{\rm BLR}^{\rm AGN}$ & $L_{\rm p, inj}$ & $\gamma_{\rm e, min}$ & $\gamma_{\rm e, max}$ & $n_{\rm p}$ & $\gamma_{\rm p, min}$ & $\gamma_{\rm p, max}$ & $L_{\rm e, in}^{\rm k}$ & $L_{\rm p,in}^{\rm k}$ & \\
parameters & [erg/s] & [erg/s] & & & & & &[erg/s] &[erg/s] & \\
\hline
Values & $5\times10^{43}$ & $1.97\times10^{44}$ & 50 & $10^7$ & 2 & 1 & $1.09\times10^{7}$ &  $1.02\times10^{44}$ & $1.3\times10^{47}$ & \\
\hline
\end{tabular}
\end{table*}

We next consider a case where the blob radius is related to the blob distance from the SMBH. In particular, we assume
that both blobs occupy the entire cross section of a conical jet, with $R_{\rm out}$ and $R_{\rm in}$ following the relation $R_{\rm out}/r_{\rm out}=R_{\rm in}/r_{\rm in}$. We also consider that the blob magnetic field strength decreases along the jet in the same way as the poloidal component of the jet's magnetic field in magnetohydrodynamical models (i.e., $B\propto r^{-1}$) \citep[e.g., ][]{2004Ap&SS.293...67V, 2007MNRAS.380...51K, 2009MNRAS.400...26O,2014Natur.510..126Z, 2018ApJ...859..168Y}.
In addition, we relax the assumption of the same particle injection spectrum for the two blobs. The fitting result is shown in the right panel of Fig.~\ref{fig:alt_fit} and the free parameters are listed in Table~\ref{tab3}.

\begin{table*}
\caption{Model parameters for the SED fit shown in the right panel of Fig.~\ref{fig:alt_fit}.}\label{tab3}
\centering
\begin{tabular}{@{}llllllllllll@{}}
\hline\hline
Free parameters&$\delta_{\rm D}$& $B_{\rm in}$&$R_{\rm in}$&$L_{\rm e, inj, in}$ &   $r^{\rm AGN}_{\rm in}$   & $n_{\rm e, 1, in}$& $n_{\rm e, 2, in}$ &$\gamma_{\rm e, b, in}$\\
 & &  $B_{\rm out}$  & $R_{\rm out}$ & $L_{\rm e, inj, out}$&$r^{\rm AGN}_{\rm out}$ & $n_{\rm e, 1, out}$ & $n_{\rm e, 2, out}$ & $\gamma_{\rm e, b, out}$  \\
 & &  [G] &    [cm] &  [erg/s] & [pc]  & & & \\
\hline
Values&26.5&0.37& $2\times10^{16}$  & $1.12\times10^{42}$ &0.08  &2  & 3.8& $1\times10^{4}$  \\
  & & 0.006 & $1.25\times10^{18}$  & $5.7\times10^{43}$ &5 & 2.2 & 4.5 & $5.93\times10^{4}$\\
\hline\hline
Fixed/Derived &  $L_{\rm BLR}^{\rm AGN}$ &  $L_{\rm p, inj}$ & $\gamma_{\rm e, min}$ & $\gamma_{\rm e, max}$ & $n_{\rm p}$ & $\gamma_{\rm p, min}$ & $\gamma_{\rm p, max}$ & $L_{\rm e, in}^{\rm k}$ & $L_{\rm p,in}^{\rm k}$\\
parameters & [erg/s] & [erg/s]  & & & & & &[erg/s] &[erg/s] \\
\hline
Values & $5\times10^{43}$ & $1.97\times10^{44}$ & 50 & $10^7$ & 2 & 1 & $4.74\times10^{7}$ & $1.24\times10^{44}$ &$1.3\times10^{47}$ \\
\hline
\end{tabular}
\end{table*}

\section{Contribution of the outer blob's radiation to the opacity of gamma-ray photons emitted by the inner blob}\label{appendix:opa}
\begin{figure}[htbp]
\centering
\includegraphics[width=0.5\textwidth]{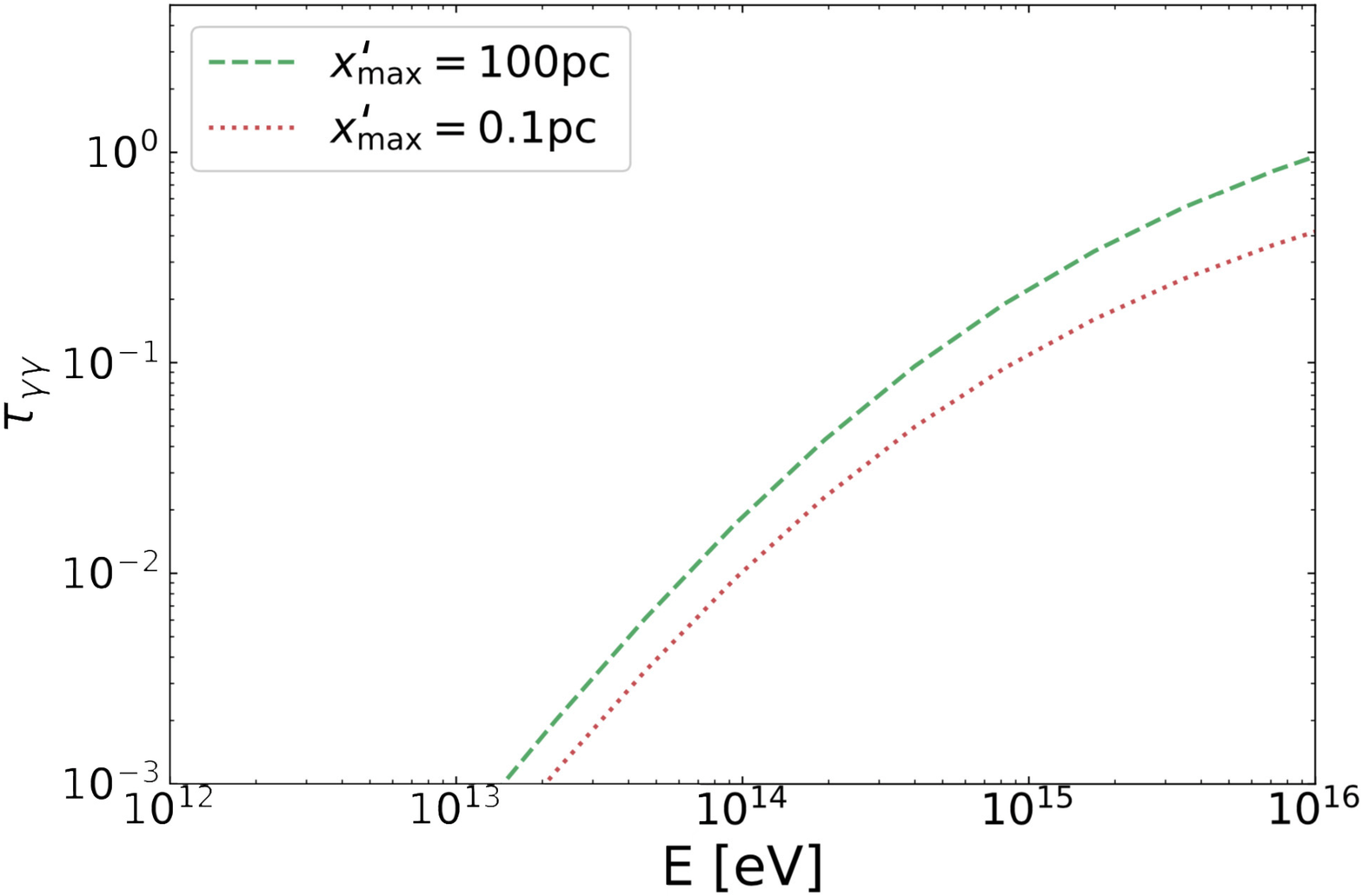}
\caption{$\gamma\gamma$ absorption opacity due to the radiation of the outer blob for high-energy photons emitted from the inner blob. The horizontal axis is the gamma-ray energy in the AGN frame. The dashed green curve and the dotted red curve show the opacity after high-energy photons having travelled a distance of 100\,pc and 0.1\,pc, respectively, in the comoving frame. \label{fig:opa_num}}
\end{figure}

In the SED fitting in the main text (Fig.~\ref{fig:sed}), we employ $\delta_{\rm D}=\Gamma=26.5$ which corresponds to a viewing angle of about 2 degree with respect to the jet's axis. In other word, we can observe photons/neutrinos which are emitted into this specific direction (in the AGN frame). The gamma-ray photons emitted from the inner blob may encounter the photon radiated by the outer blob and might be absorbed via $\gamma\gamma$ annihilation. Let us calculate the opacity of this process in the jet comoving frame and hereafter we denote quantities in jet's comoving frame with primes. Since we consider blobs as stationary features (see Section~\ref{sec:modeling} in the jet, the distance between the two blobs in the jet's comoving frame becomes $d/\delta_D$. According to the Lorentz transform of the cosine of an angle, a photon that is emitted towards observer's direction has an angle $\theta' \sim90$ degree with respect to the jet axis in the jet's comoving frame (i.e., $\mu'=(\mu-\beta)/(1-\mu \beta)$, where $\mu=\cos\theta$ and $\mu'=\cos\theta'$, $\beta$ is the jet velocity in unit of $c$). Noting that the outer blob's emission is isotropic in the jet's comoving frame and regarding the outer blob as a point source, we can integrate over the path of the gamma-ray photon to get the opacity with a delta-function approximation for the cross section of $\gamma\gamma$ annihilation, i.e.,
\begin{equation}
\tau_{\gamma\gamma}=\int_0^\infty\frac{\sigma_{\gamma\gamma}(1-\cos\alpha')L_{\rm syn}/\delta_{\rm D}^4}{4\pi \left[ (d/\delta_{\rm D})^2+x'^2\right]c(\epsilon_{\rm opt}/\delta_{\rm D}) }dx'=\frac{\sigma_{\gamma\gamma}L_{\rm syn}}{4\pi d\delta_{\rm D}^2c\epsilon_{\rm syn}}\int_0^\infty\frac{1-y/(1+y^2)^{1/2}}{1+y^2}dy=\frac{(\pi/2-1)\sigma_{\gamma\gamma}L_{\rm syn}}{4\pi d\delta_{\rm D}^2c\epsilon_{\rm syn}},
\end{equation}
where $y\equiv x'\delta_{\rm D}/d$ with $x'$ denoting the distance travelled by the high-energy photon in the jet's comoving frame. $\cos\alpha' =  x'/[x'^2+(d/\delta_{\rm D})^2]^{1/2}$ is the cosine of the collision angle between the gamma-ray photon and the soft photon from the outer blob,  $L_{\rm syn}$ is the isotropic-equivalent luminosity of synchrotron radiation of the outer blob and $\epsilon_{\rm syn}$ is the peak energy of the synchrotron radiation (in AGN frame). We are concerned with the opacity for high-energy photons co-produced with neutrinos (e.g., 1\,PeV photons), thus we look for parameters of soft photon at $\epsilon_{\rm syn}=10^{12}{\rm \,eV^2}\delta_{\rm D}^2/10^{15}{\rm \,eV}=0.7\rm \,eV$, where $L_{\rm syn}=10^{46}$ erg/s is employed. For $d=10\,$pc, we have $\tau_{\gamma\gamma}(\rm \,1PeV)=0.06$ which can be neglected. We verify the above analysis in Fig.~\ref{fig:opa_num} by calculating the opacity numerically with the synchrotron radiation spectrum obtained in Fig.~\ref{fig:sed} and the full cross section of $\gamma\gamma$ annihilation \citep{1983Afz....19..323A}. We can see that the numerical result at 1\,PeV is only a factor of 2 larger than the analytical estimation when integrating $x'$ up to 100\,pc, which is still smaller than unity for photon of energy $<$10\,PeV. We also show the result with integrating $x'$ up to 0.1\,pc for comparison. Therefore, we conclude that the radiation from the outer blob has little influence on the gamma-ray photons emitted by the inner blob in our model (at least with the employed parameters).


\begin{figure}[htbp]
\centering
\includegraphics[width=0.9\textwidth]{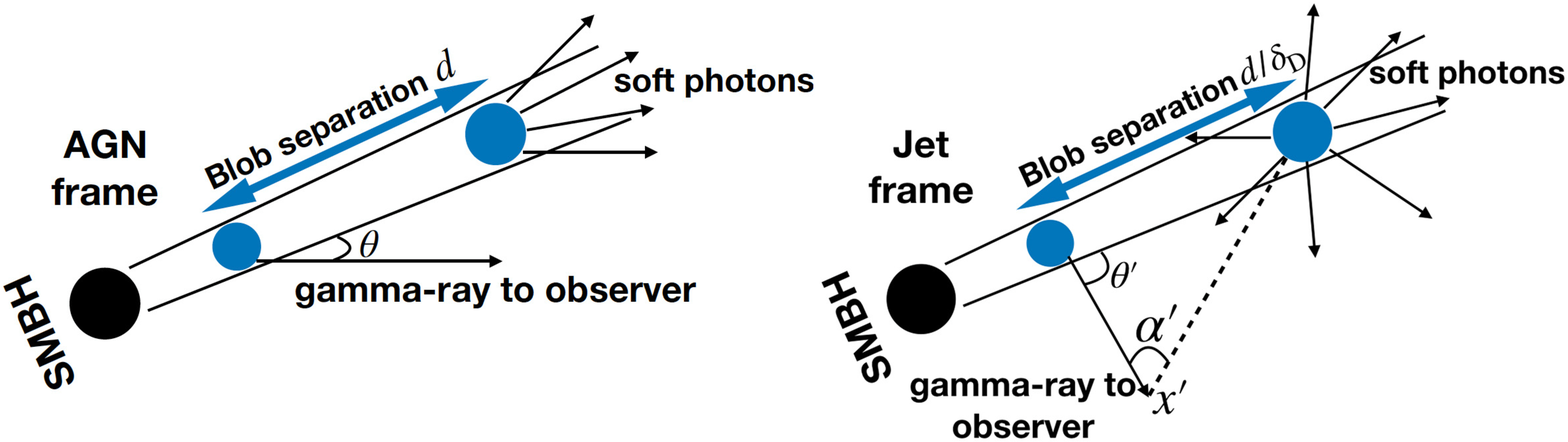}
\caption{A sketch, not to scale, of the radiation direction in different reference frames. The left and the right panel shows the geometry in the AGN frame and the jet's comoving frame respectively. A high-energy photon that propagates towards the observer moves at an angle $\theta'$ with respect to the jet's axis. The high-energy photon collides with a low-energy photon from the outer blob at a distance of $x'$ from the inner blob with a collision angle $\alpha'$. See Appendix~\ref{appendix:opa} for more discussion.\label{fig:opa} }
\end{figure}

\begin{thebibliography}{}
\bibitem[Aartsen et al.(2017)]{2017ApJ...835...45A} Aartsen, M.~G., Abraham, K., Ackermann, M., et al.\ 2017, \apj, 835, 45 

\bibitem[Abdo et al.(2010)]{2010ApJ...716...30A} Abdo, A.~A., Ackermann, M., Agudo, I., et al.\ 2010, \apj, 716, 30 

\bibitem[Aharonian et al.(1983)]{1983Afz....19..323A} Aharonian, F.~A., Atoian, A.~M., \& Nagapetian, A.~M.\ 1983, Astrofizika, 19, 323

\bibitem[Ajello et al.(2014)]{2014ApJ...780...73A} Ajello, M., Romani, R.~W., Gasparrini, D., et al.\ 2014, \apj, 780, 73 

\bibitem[Alvarez-Mu{\~n}iz, \& M{\'e}sz{\'a}ros(2004)]{2004PhRvD..70l3001A} Alvarez-Mu{\~n}iz, J., \& M{\'e}sz{\'a}ros, P.\ 2004, Physical Review D, 70, 123001

\bibitem[Ansoldi et al.(2018)]{2018ApJ...863L..10A} {Ansoldi}, S., et al. \ 2018, \apjl, 862, L10

\bibitem[Atoyan, \& Dermer(2001)]{2001PhRvL..87v1102A} Atoyan, A., \& Dermer, C.~D.\ 2001, \prl, 87, 221102


\bibitem[Atoyan \& Dermer (2003)]{2003ApJ...586...79A} Atoyan, A.~M., \& Dermer, C.~D.\ 2003, \apj, 586, 79 

\bibitem[Bednarek \& Protheroe (1999)]{1999MNRAS.302..373B} {Bednarek}, W. \& {Protheroe}, R.~J. \ 1999, \mnras, 302, 373

\bibitem[B{\"o}ttcher et al.(2013)]{2013ApJ...768...54B} B{\"o}ttcher, M., Reimer, A., Sweeney, K., \& Prakash, A.\ 2013, \apj, 768, 54 


\bibitem[Celotti \& Ghisellini(2008)]{2008MNRAS.385..283C} Celotti, A., \& Ghisellini, G.\ 2008, \mnras, 385, 283 


\bibitem[Cerruti et al.(2019)]{2019MNRAS.483L..12C} Cerruti, M., Zech, A., Boisson, C., et al.\ 2019, \mnras, 483, L12 


\bibitem[Chodorowski et al.(1992)]{1992ApJ...400..181C} Chodorowski, M.~J., Zdziarski, A.~A., \& Sikora, M.\ 1992, \apj, 400, 181


\bibitem[Cleary et al.(2007)]{2007ApJ...660..117C} Cleary, K., Lawrence, C.~R., Marshall, J.~A., Hao, L., \& Meier, D.\ 2007, \apj, 660, 117 


\bibitem[Costamante et al.(2018)]{2018MNRAS.477.4749C} Costamante, L., Cutini, S., Tosti, G., et al.\ 2018, \mnras, 477, 4749


\bibitem[Dermer et al.(2012)]{2012ApJ...755..147D} Dermer, C.~D., Murase, K., \& Takami, H.\ 2012, \apj, 755, 147 

\bibitem[Dermer et al.(2014)]{2014JHEAp...3...29D} Dermer, C.~D., Murase, K., \& Inoue, Y.\ 2014, Journal of High Energy Astrophysics, 3, 29

\bibitem[de Angelis et al.(2018)]{2018JHEAp..19....1D} de Angelis, A., Tatischeff, V., Grenier, I.~A., et al.\ 2018, Journal of High Energy Astrophysics, 19, 1

\bibitem[Dom{\'{\i}}nguez et al.(2011)]{2011MNRAS.410.2556D} Dom{\'{\i}}nguez, A., Primack, J.~R., Rosario, D.~J., et al.\ 2011, \mnras, 410, 2556 


\bibitem[Fromm et al.(2011)]{2011A&A...531A..95F} Fromm, C.~M., Perucho, M., Ros, E., et al.\ 2011, \aap, 531, A95


\bibitem[Gao et al.(2019)]{2019NatAs...3...88G} Gao, S., Fedynitch, A., Winter, W., \& Pohl, M.\ 2019, Nature Astronomy, 3, 88 


\bibitem[Georganopoulos \& Marscher(1998)]{1998ApJ...506..621G} Georganopoulos, M., \& Marscher, A.~P.\ 1998, \apj, 506, 621 


\bibitem[Ghisellini et al.(2010)]{2010MNRAS.402..497G} Ghisellini, G., Tavecchio, F., Foschini, L., et al.\ 2010, \mnras, 402, 497 


\bibitem[Giommi et al.(2012)]{2012MNRAS.420.2899G} Giommi, P., Padovani, P., Polenta, G., et al.\ 2012, \mnras, 420, 2899 


\bibitem[Giommi et al.(2013)]{2013MNRAS.431.1914G} Giommi, P., Padovani, P., \& Polenta, G.\ 2013, \mnras, 431, 1914 


\bibitem[Harris \& Krawczynski(2006)]{2006ARA&A..44..463H} Harris, D.~E., \& Krawczynski, H.\ 2006, \araa, 44, 463 


\bibitem[Hayashida et al.(2012)]{2012ApJ...754..114H} Hayashida, M., Madejski, G.~M., Nalewajko, K., et al.\ 2012, \apj, 754, 114 


\bibitem[IceCube Collaboration et al.(2018)]{2018Sci...361.1378I} IceCube Collaboration, Aartsen, M.~G., Ackermann, M., et al.\ 2018, Science, 361, eaat1378 

\bibitem[IceCube Collaboration et al.(2018b)]{2018Sci...361..147I} IceCube Collaboration, Aartsen, M.~G., Ackermann, M., et al.\ 2018, Science, 361, 147

\bibitem[Inoue et al.(2019)]{2019ApJ...880...40I} Inoue, Y., Khangulyan, D., Inoue, S., et al.\ 2019, The Astrophysical Journal, 880, 40

\bibitem[Kamraj et al.(2018)]{2018ApJ...866..124K} Kamraj, N., Harrison, F.~A., Balokovi{\'c}, M., et al.\ 2018, \apj, 866, 124

\bibitem[Katarzy{\'n}ski et al.(2001)]{2001A&A...367..809K} Katarzy{\'n}ski, K., Sol, H., \& Kus, A.\ 2001, \aap, 367, 809

\bibitem[Kimura et al.(2015)]{2015ApJ...806..159K} Kimura, S.~S., Murase, K., \& Toma, K.\ 2015, The Astrophysical Journal, 806, 159

\bibitem[Keivani et al.(2018)]{2018ApJ...864...84K} Keivani, A., Murase, K., Petropoulou, M., et al.\ 2018, \apj, 864, 84 


\bibitem[Kelner \& Aharonian(2008)]{2008PhRvD..78c4013K} Kelner, S.~R., \& Aharonian, F.~A.\ 2008, \prd, 78, 034013

\bibitem[Komissarov et al.(2007)]{2007MNRAS.380...51K} {Komissarov}, S. S., et al. \ 2007, \mnras, 380, 1

\bibitem[Lagage \& Cesarsky(1983)]{1983A&A...125..249L} Lagage, P.~O., \& Cesarsky, C.~J.\ 1983, \aap, 125, 249 


\bibitem[Liu et al.(2019)]{2019PhRvD..99f3008L} Liu, R.-Y., Wang, K., Xue, R., et al.\ 2019, \prd, 99, 063008 


\bibitem[Madejski et al.(2016)]{2016ApJ...831..142M} Madejski, G.~M., Nalewajko, K., Madsen, K.~K., et al.\ 2016, \apj, 831, 142 


\bibitem[Mannheim \& Biermann(1992)]{1992A&A...253L..21M} Mannheim, K., \& Biermann, P.~L.\ 1992, \aap, 253, L21 

\bibitem[Mannheim et al.(1992)]{1992A&A...260L...1M} Mannheim, K. Stanev, T., \& {Biermann}, P.~L. \ 1992, \aap, 260, L1

\bibitem[Mastichiadis \& Kirk (1997)]{1997A&A...320...19M}
Mastichiadis, A., \& Kirk, J.~G., 1997, \aap, 320

\bibitem[Moderski et al.(2005)]{2005MNRAS.363..954M} Moderski, R., Sikora, M., Coppi, P.~S., \& Aharonian, F.\ 2005, \mnras, 363, 954 

\bibitem[Moiseev, \& Amego Team(2017)]{2017ICRC...35..798M} Moiseev, A., \& Amego Team\ 2017, 35th International Cosmic Ray Conference (ICRC2017), 798

\bibitem[Murase et al.(2014)]{2014PhRvD..90b3007M} Murase, K., Inoue, Y., \& Dermer, C.~D.\ 2014, \prd, 90, 23007

\bibitem[Murase et al.(2016)]{2016PhRvL.116g1101M} Murase, K., Guetta, D., \& Ahlers, M.\ 2016, Physical Review Letters, 116, 071101


\bibitem[Murase et al.(2018)]{2018ApJ...865..124M} Murase, K., Oikonomou, F., \& Petropoulou, M.\ 2018, \apj, 865, 124 

\bibitem[Murase et al.(2019)]{2019arXiv190404226M} Murase, K., Kimura, S.~S., \& Meszaros, P.\ 2019, arXiv e-prints, arXiv:1904.04226

\bibitem[Nalewajko et al.(2012)]{2012ApJ...760...69N} Nalewajko, K., Sikora, M., Madejski, G.~M., et al.\ 2012, \apj, 760, 69

\bibitem[O'Sullivan, \& Gabuzda(2009)]{2009MNRAS.400...26O} O'Sullivan, S.~P., \& Gabuzda, D.~C.\ 2009, \mnras, 400, 26


\bibitem[Padovani et al.(2016)]{2016MNRAS.457.3582P} Padovani, P., Resconi, E., Giommi, P., Arsioli, B., \& Chang, Y.~L.\ 2016, \mnras, 457, 3582 


\bibitem[Padovani et al.(2019)]{2019MNRAS.484L.104P} Padovani, P., Oikonomou, F., Petropoulou, M., Giommi, P., \& Resconi, E.\ 2019, \mnras, 484, L104

\bibitem[Paiano et al.(2018)]{paiano18} Paiano S., Falomo R., Treves A., \& Scarpa R. \ 2018, \apjl, 854, L3

\bibitem[Paladino et al.(2019)]{2019ApJ...871...41P} {Palladino}, A., {Rodrigues}, X., {Gao}, S., \& {Winter}, W. \ 2019, \apj, 871, 41

\bibitem[Plavin et al.(2019)]{2019ApJ...871..143P} Plavin, A.~V., Kovalev, Y.~Y., \& Petrov, L.~Y.\ 2019, \apj, 871, 143

\bibitem[Petropoulou et al.(2015)]{2015MNRAS.448.2412P} {Petropoulou}, M., {Dimitrakoudis}, S., {Padovani}, P.,       {Mastichiadis}, A., \& {Resconi}, E. \ 2015, \mnras, 448, 2412

\bibitem[Petropoulou \& Mastichiadis (2015)]{2015MNRAS.447...36P} Petropoulou, M., \& Mastichiadis, A. \ 2015, \mnras, 447, 36 

\bibitem[Petropoulou et al.(2017)]{2017MNRAS.470.1881P} Petropoulou, M., Coenders, S., Vasilopoulos, G., Kamble, A., \& Sironi, L.\ 2017, \mnras, 470, 1881 

\bibitem[Prince et al.(2019)]{2019arXiv190804803P} Prince, R., Gupta, N., \& Nalewajko, K.\ 2019, arXiv e-prints, arXiv:1908.04803

\bibitem[Protheroe \& Clay(2004)]{2004PASA...21....1P} Protheroe, R.~J., \& Clay, R.~W.\ 2004, \pasa, 21, 1


\bibitem[Reimer et al.(2018)]{2018arXiv181205654R} Reimer, A., Boettcher, M., \& Buson, S.\ 2018, arXiv e-prints, arXiv:1812.05654


\bibitem[Rieger et al.(2007)]{2007Ap&SS.309..119R} Rieger, F.~M., Bosch-Ramon, V., \& Duffy, P.\ 2007, \apss, 309, 119

\bibitem[Ricci et al.(2018)]{2018MNRAS.480.1819R} Ricci, C., Ho, L.~C., Fabian, A.~C., et al.\ 2018, \mnras, 480, 1819 

\bibitem[Righi et al.(2019)]{2019MNRAS.483L.127R} Righi, C., Tavecchio, F., \& Inoue, S.\ 2019, \mnras, 483, L127

\bibitem[Rodrigues et al.(2019)]{2019ApJ...874L..29R} Rodrigues, X., Gao, S., Fedynitch, A., et al.\ 2019, \apjl, 874, L29


\bibitem[Sahakyan(2018)]{2018ApJ...866..109S} Sahakyan, N.\ 2018, \apj, 866, 109

\bibitem[Stecker et al.(1991)]{1991PhRvL..66.2697S} Stecker, F.~W., Done, C., Salamon, M.~H., et al.\ 1991, Physical Review Letters, 66, 2697

\bibitem[Stickel et al.(1991)]{stickel91}Stickel M. et al. \ 1991, \apj, 374, 431

\bibitem[Stocke et al.(1991)]{stocke91} Stocke J.~T. et al. \ 1991, \apjs, 76, 8

\bibitem[Strotjohann et al.(2019)]{2019A&A...622L...9S} Strotjohann, N.~L., Kowalski, M., \& Franckowiak, A.\ 2019, \aap, 622, L9 

\bibitem[Tavecchio \& Ghisellini(2008)]{2008MNRAS.386..945T} Tavecchio, F., \& Ghisellini, G.\ 2008, \mnras, 386, 945


\bibitem[Tavecchio et al.(2011)]{2011A&A...534A..86T} Tavecchio, F., Becerra-Gonzalez, J., Ghisellini, G., et al.\ 2011, \aap, 534, A86

\bibitem[Tavecchio et al.(2014)]{2014ApJ...793L..18T} Tavecchio, F., Ghisellini, G., \& Guetta, D.\ 2014, The Astrophysical Journal, 793, L18

\bibitem[Tavecchio, \& Ghisellini(2015)]{2015MNRAS.451.1502T} Tavecchio, F., \& Ghisellini, G.\ 2015, \mnras, 451, 1502

\bibitem[Urry \& Padovani(1995)]{1995PASP..107..803U} Urry, C.~M., \& Padovani, P.\ 1995, \pasp, 107, 803 

\bibitem[Vlahakis (2004)]{2004Ap&SS.293...67V} Vlahakis, N. \ 2004, \apss, 293, 67

\bibitem[Wang et al.(2018)]{2018ApJ...857...24W} Wang, K., Liu, R.-Y., Dai, Z.-G., \& Asano, K.\ 2018, \apj, 857, 24

\bibitem[Wang et al.(2018b)]{2018arXiv180900601W} Wang, K., Liu, R.-Y., Li, Z., et al.\ 2018, arXiv e-prints, arXiv:1809.00601

\bibitem[Wu et al.(2016)]{2016SPIE.9905E..6EW} Wu, X., Walter, R., Su, M., et al.\ 2016, \procspie, 99056E

\bibitem[Xue et al.(2019)]{2019ApJ...871...81X} Xue, R., Liu, R.-Y., Wang, X.-Y., Yan, H., \& B{\"o}ttcher, M.\ 2019, \apj, 871, 81 


\bibitem[Yan \& Zhang(2015)]{2015MNRAS.447.2810Y} Yan, D., \& Zhang, L.\ 2015, \mnras, 447, 2810


\bibitem[Yan et al.(2018)]{2018ApJ...859..168Y} Yan, D., Wu, Q., Fan, X., et al.\ 2018, \apj, 859, 168


\bibitem[Zamaninasab et al.(2014)]{2014Natur.510..126Z} Zamaninasab, M., Clausen-Brown, E., Savolainen, T., et al.\ 2014, \nat, 510, 126

\bibitem[Zhang et al.(2019)]{2019ApJ...876..109Z} Zhang, H., Fang, K., Li, H., et al.\ 2019, \apj, 876, 109

\end{thebibliography}
\end{document}